\documentclass[conference]{IEEEtran}
\IEEEoverridecommandlockouts

\usepackage{amsmath,amssymb,amsthm}
\usepackage{array}
\usepackage{balance}
\usepackage{bm}
\usepackage{booktabs}
\usepackage{cite}
\usepackage{colortbl}
\usepackage{enumitem}
\usepackage{graphicx}
\usepackage{multirow}
\usepackage{xcolor}
\usepackage{subfigure}
\usepackage{tabularx}
\usepackage{textcomp}
\usepackage{threeparttable}
\usepackage[colorlinks,linkcolor=blue,citecolor=blue,urlcolor=blue]{hyperref}

\newtheoremstyle{mydefn}
{}{}
{\it}       
{0pt}       
{\bfseries} 
{:~}        
{0pt}       
{}          
\theoremstyle{mydefn}

\newtheorem{definition}{Definition}
\newtheorem{lemma}{Lemma}
\newtheorem{theorem}{Theorem}
\newtheorem{corollary}{Corollary}

\newtheoremstyle{myexample}
{}{}
{}          
{0pt}       
{\bfseries} 
{:~}        
{0pt}       
{}          
\theoremstyle{myexample}

\newtheorem{example}{Example}

\renewenvironment{proof}{\smallskip\noindent{\bfseries Proof:}}{\qed \smallskip}

\renewcommand{\paragraph}[1]{\smallskip\noindent\textbf{#1.}}
\newcommand{\subparagraph}[1]{\smallskip\noindent\textit{\underline{#1.}}}

\DeclareMathOperator*{\argmax}{arg\,max}

\newcommand{\norm}[1]{\Vert #1 \Vert}
\newcommand{\num}[1]{\vert #1 \vert}
\newcommand{\abso}[1]{\vert #1 \vert}
\newcommand{\ip}[1]{\langle #1 \rangle}

\makeatletter
\newif\if@restonecol
\makeatother

\usepackage[linesnumbered,ruled,vlined]{algorithm2e}
\usepackage{algpseudocode}

\SetKwComment{Comment}{$\triangleright$\ }{}

\def\BibTeX{{\rm B\kern-.05em{\sc i\kern-.025em b}\kern-.08em
    T\kern-.1667em\lower.7ex\hbox{E}\kern-.125emX}}

\begin{document}

\title{Diversity-Aware $k$-Maximum Inner Product Search Revisited}

\author{
\IEEEauthorblockN{
  Qiang Huang,\IEEEauthorrefmark{2}
  Yanhao Wang,\IEEEauthorrefmark{4}
  Yiqun Sun,\IEEEauthorrefmark{2}
  Anthony K. H. Tung\IEEEauthorrefmark{2}}  
\IEEEauthorblockA{
  \IEEEauthorrefmark{2}\textit{School of Computing, National University of Singapore, Singapore}}
\IEEEauthorblockA{
  \IEEEauthorrefmark{4}\textit{School of Data Science and Engineering, East China Normal University, Shanghai, China}}
\IEEEauthorblockA{
  huangq@comp.nus.edu.sg, 
  yhwang@dase.ecnu.edu.cn,
  sunyq@comp.nus.edu.sg, 
  atung@comp.nus.edu.sg}
}

\maketitle

\begin{abstract}
The $k$-Maximum Inner Product Search ($k$MIPS) serves as a foundational component in recommender systems and various data mining tasks. However, while most existing $k$MIPS approaches prioritize the efficient retrieval of highly relevant items for users, they often neglect an equally pivotal facet of search results: \emph{diversity}.
To bridge this gap, we revisit and refine the diversity-aware $k$MIPS (D$k$MIPS) problem by incorporating two well-known diversity objectives -- minimizing the average and maximum pairwise item similarities within the results -- into the original relevance objective. This enhancement, inspired by Maximal Marginal Relevance (MMR), offers users a controllable trade-off between relevance and diversity.
We introduce \textsc{Greedy} and \textsc{DualGreedy}, two linear scan-based algorithms tailored for D$k$MIPS. They both achieve data-dependent approximations and, when aiming to minimize the average pairwise similarity, \textsc{DualGreedy} attains an approximation ratio of $1/4$ with an additive term for regularization. To further improve query efficiency, we integrate a lightweight Ball-Cone Tree (BC-Tree) index with the two algorithms.
Finally, comprehensive experiments on ten real-world data sets demonstrate the efficacy of our proposed methods, showcasing their capability to efficiently deliver diverse and relevant search results to users. 
\end{abstract}


\section{Introduction}
\label{sect:intro}

Recommender systems have become indispensable in many real-world applications such as e-commerce \cite{smith2017two, pfadler2020billion}, streaming media \cite{covington2016deep, steck2021deep}, and news feed \cite{karimi2018news, raza2022news}.
Matrix Factorization (MF) is a dominant approach in the field of recommendation \cite{KorenBV09, XueDZHC17, rendle2020neural, anelli2021reenvisioning}.
Within the MF framework, items and users are represented as vectors in a $d$-dimensional inner product space $\mathbb{R}^d$ derived from a user-item rating matrix.
The relevance between an item and a user is assessed by the inner product of their corresponding vectors.
As such, the $k$-Maximum Inner Product Search ($k$MIPS), which identifies the top-$k$ item vectors with the largest inner products for a query (user) vector, plays a vital role in MF-based recommendation.
To enable real-time interactions with users, a substantial body of work has arisen to optimize the efficiency and scalability of $k$MIPS \cite{RamG12, Shrivastava014, TeflioudiGM15, LiCYM17, HuangMFFT18, MorozovB18, AbuzaidSBZ19, SongGZ021, TanXZFZL21, ZhaoZYLXZJ23}.

\begin{figure*}[t]%
\centering%
\subfigure[Top-10 recommendation lists.]{%
  \label{fig:motivation:posters}%
  \includegraphics[width=0.60\textwidth]{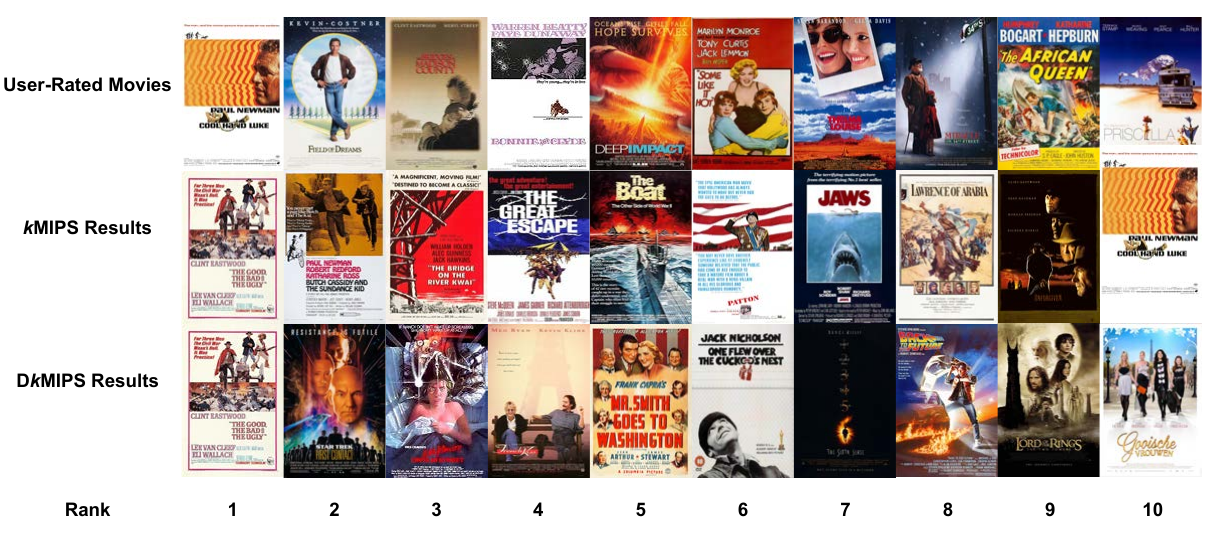}}%
\hspace{0.3em}%
\subfigure[Histograms for genre frequency.]{%
  \label{fig:motivation:hist}%
  \includegraphics[width=0.375\textwidth]{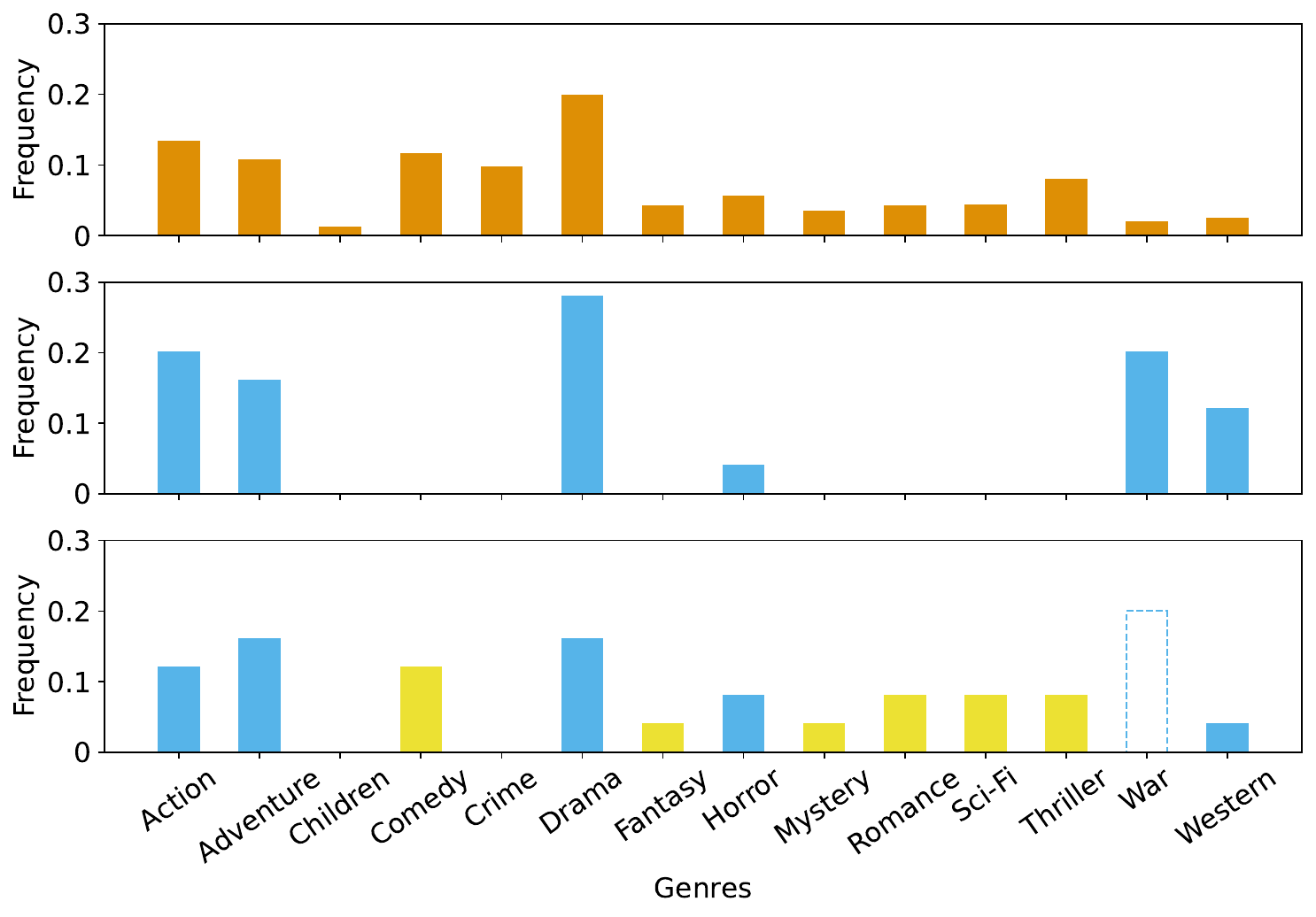}}%
\vspace{-0.5em}%
\caption{Comparison of the results provided by $k$MIPS and D$k$MIPS on the MovieLens data set when $k = 10$. In Fig.~\ref{fig:motivation:posters}, we display posters of ten randomly user-rated movies, together with those returned by $k$MIPS (using \textsc{Linear}) and D$k$MIPS (using \textsc{DualGreedy-Avg}). In Fig.~\ref{fig:motivation:hist}, we present histograms showing the genre distribution of all user-rated movies and those returned by both methods. See Section \ref{sect:expt:case_study} for detailed results and analyses.}%
\label{fig:motivation}%
\vspace{-1.0em}%
\end{figure*}

While \emph{relevance} is pivotal in enhancing recommendation quality, the importance of \emph{diversity} also cannot be overlooked \cite{AgrawalGHI09, kaminskas2016diversity, kunaver2017diversity, abdool2020managing, anderson2020algorithmic}.
Diverse results offer two compelling benefits:
(1) they mitigate the risk of overwhelming users with excessively homogeneous suggestions, thereby sustaining their engagement;
(2) they guide users across various appealing domains, aiding them to uncover untapped areas of interest.
Despite the clear benefits of diversity, existing $k$MIPS methods have focused primarily on providing relevant items to users while paying limited attention to diversity.
This issue is vividly evident in Fig.~\ref{fig:motivation}, where we present the results of an exact $k$MIPS method (by evaluating the inner products of all item vectors and the query vector using a linear scan, or \textsc{Linear}) on the MovieLens data set \cite{harper2015movielens}.
Although the user exhibits a diverse interest spanning ten different movie genres, the $k$MIPS results are predominantly limited to a narrow subset of genres: Action, Adventure, Drama, War, and Western.
This example underscores that while $k$MIPS excels in providing relevant items, it often fails to diversify the result.

As a pioneering effort, Hirata et al. \cite{hirata2022solving} first investigated the problem of diversity-aware $k$MIPS (D$k$MIPS).
This problem refines the conventional $k$MIPS by incorporating a diversity term into its relevance-based objective function.
For a given query vector $\bm{q}$, the D$k$MIPS aims to find a subset $\mathcal{S}$ of $k$ item vectors that simultaneously satisfy two criteria:
(1) each item vector $\bm{p} \in \mathcal{S}$ should have a large inner product with $\bm{q}$, signifying its relevance;
(2) the items within $\mathcal{S}$ should be distinct from each other, ensuring diversity.
Although the user-item relevance remains to be evaluated based on the inner product of their corresponding vectors, the item-wise dissimilarity is measured using a different vector representation in the Euclidean space, constructed by Item2Vec \cite{barkan2016item2vec}.
The D$k$MIPS problem is NP-hard, stemming from its connection to Maximal Marginal Relevance (MMR) \cite{carbonell1998use}, a well-recognized NP-hard problem.
Given the NP-hardness of D$k$MIPS, Hirata et al. \cite{hirata2022solving} introduced a heuristic \textsc{IP-Greedy} algorithm, which follows a greedy framework and employs pruning and skipping techniques to efficiently obtain D$k$MIPS results.

However, upon a closer examination, we uncover several limitations inherent to the D$k$MIPS problem and the \textsc{IP-Greedy} algorithm as presented in \cite{hirata2022solving}.
\begin{enumerate}
  \item The D$k$MIPS problem operates in two distinct spaces, although the item vectors of MF and Item2Vec are derived from the same rating matrix. This dual-space operation incurs extra pre-processing costs without significantly contributing new insights to enhance D$k$MIPS results.
  \item The computation of the marginal gain function in \textsc{IP-Greedy} does not align seamlessly with its defined objective function. We will delve into this misalignment more thoroughly in Section \ref{sect:problem}.
  \item \textsc{IP-Greedy} does not provide any approximation bound for the D$k$MIPS problem. Consequently, its result can be arbitrarily bad in the worst case.
  \item Despite the use of pruning and skipping techniques, \textsc{IP-Greedy} remains a linear scan-based method, which has a worst-case time complexity of $O(n d k^2 \log{n})$. This makes it susceptible to performance degradation, especially when the result size $k$ is large.
\end{enumerate}
These limitations often lead to the suboptimal recommendation quality and inferior search efficiency of \textsc{IP-Greedy}, as confirmed by our experimental results in Section \ref{sect:expt}.

\paragraph{Our Contributions}
In this paper, we revisit and improve the problem formulation and algorithms for D$k$MIPS.
Inspired by the well-established concept of MMR \cite{carbonell1998use} and similarly to \cite{hirata2022solving}, our revised D$k$MIPS objective function also takes the form of a linear combination of relevance and diversity terms.
However, we argue that item vectors derived from MF in an inner product space are already capable of measuring item similarities, and introducing an additional item representation for similarity computation offers limited benefits.
We use two common measures to evaluate how diverse a result set is, that is, the average and maximum pairwise similarity among item vectors, as computed based on their inner products.
Accordingly, our D$k$MIPS problem selects a set of $k$ items that exhibit large inner products with the query for \emph{relevance} while also having small (average or maximum) inner products with each other for \emph{diversity}.
In addition, we introduce a \emph{user-controllable} balancing parameter $\lambda \in [0, 1]$ to adjust the relative importance of these two terms.

We propose two linear scan-based approaches to answering D$k$MIPS.
The first method is \textsc{Greedy}, which runs in $k$ rounds and adds the item that can maximally increase the objective function to the result in each round.
\textsc{Greedy} achieves a data-dependent approximation factor for D$k$MIPS in $O(ndk^2)$ time, regardless of the diversity measure used.
Our second method, \textsc{DualGreedy}, which maintains two results greedily in parallel and returns the better one between them for D$k$MIPS, is particularly effective when using the average of pairwise inner products to measure diversity.
It takes advantage of the submodularity of the objective function \cite{KrauseG14, HanCCW20}, achieving a data-independent approximation factor of $1/4$ with an additive regularization term, also in $O(ndk^2)$ time.
Fig.~\ref{fig:motivation} illustrates that \textsc{DualGreedy-Avg}, which measures diversity using the average (\textsc{Avg}) of pairwise inner products, yields more diverse query results, covering a wider range of user-preferred genres like Comedy, Romance, Sci-Fi, and Thriller. 
Moreover, we develop optimization techniques that reduce their time complexity to $O(ndk)$, significantly improving efficiency.

To further elevate the real-time query processing capability of \textsc{Greedy} and \textsc{DualGreedy}, we integrate an advanced Ball-Cone Tree (BC-Tree) structure \cite{huang2023lightweight} into both algorithms, resulting in \textsc{BC-Greedy} and \textsc{BC-DualGreedy}, respectively.
This integration utilizes the ball and cone structures in BC-Tree for pruning, which expedites the identification of the item with the maximal marginal gain at each iteration.
Note that we do not employ other prevalent data structures, such as locality-sensitive hashing (LSH) \cite{Shrivastava014, HuangMFFT18, ZhaoZYLXZJ23}, quantization \cite{GuoKCS16, Zhang_Lian_Zhang_Wang_Chen_2023}, and proximity graphs \cite{MorozovB18, TanXZFZL21}, as the marginal gain function involves two distinct terms for relevance and diversity. This dual-term nature in our context complicates meeting the fundamental requirements of these structures.

Finally, we conduct extensive experiments on ten public real-world data sets to thoroughly evaluate our proposed techniques.
The results underscore their exceptional efficacy.
Specifically, \textsc{Greedy} and \textsc{DualGreedy} consistently outperform \textsc{IP-Greedy} and \textsc{Linear}, providing users with recommendations that are much more diverse but equally relevant.
Furthermore, by leveraging the BC-Tree structure, \textsc{BC-Greedy} and \textsc{BC-DualGreedy} not only produce high-quality recommendations but also stand out as appropriate solutions in real-time recommendation contexts.

\paragraph{Paper Organization}
The remainder of this paper is organized as follows.
The basic notions and problem formulation are given in Section \ref{sect:problem}.
Section \ref{sect:methods} presents \textsc{Greedy} and \textsc{DualGreedy}, which are further optimized with BC-Tree integration in Section \ref{sect:bctree}. 
The experimental results and analyses are provided in Section \ref{sect:expt}.
Section \ref{sect:related_work} reviews the related work.
We conclude this paper in Section \ref{sect:conclusions}.

\section{Problem Formulation}
\label{sect:problem}

\noindent\textbf{$k$MIPS and D$k$MIPS.}
In this paper, we denote $\mathcal{P}$ as a data set of $n$ item vectors in a $d$-dimensional inner product space $\mathbb{R}^d$, and $\mathcal{Q}$ as a user set of $m$ user vectors in the same space. Each query vector $\bm{q}=(q_1,\cdots,q_d)$ is drawn uniformly at random from $\mathcal{Q}$.
The inner product between any pair of vectors $\bm{p}$ and $\bm{q}$ in $\mathbb{R}^d$ is represented as $\ip{\bm{p},\bm{q}} = \sum_{i=1}^{d} p_i q_i$.
We define the $k$MIPS problem as follows:
\vspace{-0.25em}
\begin{definition}[$k$MIPS]\label{def:kmips}
  Given a set of $n$ item vectors $\mathcal{P} \subset \mathbb{R}^{d}$, a query vector $\bm{q} \in \mathbb{R}^d$, and an integer $k \geq 1$, identify a set of $k$ item vectors $\mathcal{S} \subseteq \mathcal{P}$ s.t. $\ip{\bm{p}, \bm{q}} \geq \ip{\bm{p}^\prime, \bm{q}}$ for every $\bm{p} \in \mathcal{S}$ and $\bm{p}^\prime \in \mathcal{P}\setminus\mathcal{S}$ (ties are broken arbitrarily). 
\end{definition}

We revisit the D$k$MIPS problem in \cite{hirata2022solving}, where each item is denoted by two distinct vectors: (1) $\bm{p}$ in the inner product space generated by MF to assess item-user relevance; (2) $\hat{\bm{p}}$ in the Euclidean space generated by Item2Vec \cite{barkan2016item2vec} to measure item dissimilarity. The D$k$MIPS problem is defined as:
\vspace{-0.25em}
\begin{definition}[D$k$MIPS \cite{hirata2022solving}]\label{def:orig-diverse-kmips}
  Given a set of $n$ items $\mathcal{P} \subset \mathbb{R}^{d}$ with each represented as two vectors $\bm{p}$ and $\hat{\bm{p}}$, a query vector $\bm{q} \in \mathbb{R}^d$, an integer $k > 1$, a balancing factor $\lambda \in [0,1]$, and a scaling factor $\mu > 0$, find a set $\mathcal{S}^*$ of $k$ item vectors s.t.
  \vspace{-0.15em}
  \begin{equation}\label{eqn:optimal-S}
    \mathcal{S}^* = \textstyle \argmax_{\mathcal{S} \subset \mathcal{P}, \num{\mathcal{S}}=k} f(\mathcal{S}),
  \end{equation}
  where the objective function $f(\mathcal{S})$ is defined as:
  \vspace{-0.15em}
  \begin{equation}\label{eqn:f_S} 
    f(\mathcal{S}) = \tfrac{\lambda}{k} \textstyle \sum_{\bm{p}\in \mathcal{S}} \ip{\bm{p}, \bm{q}} + \mu(1-\lambda) \textstyle \min_{\hat{\bm{p}} \neq \hat{\bm{p}}^\prime \in \mathcal{S}} \norm{\hat{\bm{p}} - \hat{\bm{p}}^\prime}.
  \end{equation}
\end{definition}

\noindent\textbf{Misalignment in \textsc{IP-Greedy}.}
Based on Definition \ref{def:orig-diverse-kmips}, \textsc{IP-Greedy} \cite{hirata2022solving} iteratively identifies the item $\bm{p}$ maximizing $f(\mathcal{S})$ w.r.t.~the current result set $\mathcal{S}$. 
The marginal gain function of adding $\bm{p}$ into $\mathcal{S}$ is defined below:
\vspace{-0.15em}
\begin{equation}\label{eqn:incremental-f_S}
  \resizebox{0.91\hsize}{!}{%
  $\Delta_f(\bm{p},\mathcal{S}) = \lambda \ip{\bm{p},\bm{q}} + \mu(1-\lambda) \textstyle \min_{\hat{\bm{p}}_x \neq \hat{\bm{p}}_y \in \mathcal{S} \cup \{\hat{\bm{p}}\}} \norm{\hat{\bm{p}}_x - \hat{\bm{p}}_y}.$
  }
\end{equation}

However, we find that the sum of $\Delta_f(\bm{p},\mathcal{S})$ from the $k$ items identified by \textsc{IP-Greedy} does not align with the $f(\mathcal{S})$ defined by Eq.~\ref{eqn:f_S}. 
This discrepancy arises for two reasons.
First, the scaling factor in the first term in Eq.~\ref{eqn:incremental-f_S} is $\lambda$, as opposed to $\tfrac{\lambda}{k}$ used in $f(\mathcal{S})$.
More importantly, the second term in Eq.~\ref{eqn:incremental-f_S} computes the minimum Euclidean distance among all items found so far, rather than the incremental change of the second term in $f(\mathcal{S})$ when adding $\bm{p}$.
This leads to the minimum distance being counted multiple times (e.g., $k$ times) for possibly different pairs of items.
Since the items contributing to the minimal distance often \emph{vary} over iterations, the result set $\mathcal{S}$ may not truly represent $f(\mathcal{S})$, leading to significant performance degradation. 
We will substantiate this using experimental results in Section \ref{sect:expt:recommendation}.

\begin{table}[t]
\centering
\footnotesize
\caption{List of frequently used notations.}
\vspace{-0.5em}
\label{tab:notations}
\resizebox{.99\columnwidth}{!}{%
  \begin{tabular}{ll} \toprule
    \textbf{Symbol} & \textbf{Description} \\
    \midrule
    $\mathcal{P}$, $n$ & A data set $\mathcal{P}$ of $n$ items, i.e., $\mathcal{P} \subseteq \mathbb{R}^d$ and $\num{\mathcal{P}} = n$ \\
    $\mathcal{S}$, $k$ & A result set $\mathcal{S}$ of $k$ items, i.e., $\mathcal{S} \subseteq \mathcal{P}$ and $\num{\mathcal{S}} = k$ \\
    $\bm{p}$, $\bm{q}$ & An item vector and a query (user) vector \\
    $\lambda$, $\mu$ & A balancing factor $\lambda \in [0,1]$ and a scaling factor $\mu > 0$ \\
    $f(\mathcal{S})$ & The D$k$MIPS objective function for the result set $\mathcal{S}$ \\
    $\Delta_{f}(\bm{p},\mathcal{S})$ & The marginal gain of $\bm{p}$ w.r.t.~$\mathcal{S}$ \\
    $\ip{\cdot,\cdot}$ & The inner product of any two vectors \\
    $\norm{\cdot}$ & The $l_2$-norm of a vector (or $l_2$-distance of two vectors) \\
    $\mathcal{N}$ & An (internal or leaf) node of the BC-Tree \\
    $\mathcal{N}.\mathcal{L}$, $\mathcal{N}.\mathcal{R}$ & The left and right children of a node $\mathcal{N}$ \\
    $\mathcal{N}.\bm{c}$, $\mathcal{N}.r$  & The center and radius of a node $\mathcal{N}$ \\
    $r_{\bm{p}}$, $\varphi_{\bm{p}}$ & The radius and angle between $\bm{p}$ and the leaf center $\mathcal{N}.\bm{c}$ \\
    \bottomrule 
  \end{tabular}
}
\end{table}
\setlength{\textfloatsep}{1.0em}

\paragraph{New Formulation of D$k$MIPS}
In this work, we simplify the pre-processing stage by focusing on a single space and propose a new D$k$MIPS formulation. 
We assess the \emph{diversity} of a result set $\mathcal{S}$ by analyzing the inner products among its item vectors. Notably, a larger inner product indicates higher similarity between items, which in turn signifies reduced diversity.
We use two standard measures to quantify diversity: the average and maximum pairwise inner products among the items in $\mathcal{S}$, to provide a comprehensive assessment of diversity:
\begin{definition}[D$k$MIPS, revisited]\label{def:diverse-kmips}
  Suppose that we utilize the same inputs, notations, and Eq.~\ref{eqn:optimal-S} as specified in Definition \ref{def:orig-diverse-kmips} excluding the vector $\hat{\bm{p}}$ for each item.
  We focus on the objective function $f(\mathcal{S})$, which is defined by Eq.~\ref{eqn:f_S-avg} or~\ref{eqn:f_S-max} below, each of which employs one of the two diversity measures for $\mathcal{S}$:
  \begin{align}
    & f_{avg}(\mathcal{S}) = \tfrac{\lambda}{k} \textstyle \sum_{\bm{p}\in \mathcal{S}} \ip{\bm{p}, \bm{q}} - \tfrac{2\mu(1-\lambda)}{k(k-1)} \textstyle \sum_{\bm{p} \neq \bm{p}^\prime \in \mathcal{S}} \ip{\bm{p}, \bm{p}^\prime}; \label{eqn:f_S-avg} \\
    &
    \resizebox{0.91\hsize}{!}{%
    $f_{max}(\mathcal{S}) = \tfrac{\lambda}{k} \textstyle \sum_{\bm{p}\in \mathcal{S}} \ip{\bm{p}, \bm{q}} - \mu(1-\lambda) \max_{\bm{p} \neq \bm{p}^\prime \in \mathcal{S}} \ip{\bm{p}, \bm{p}^\prime}.$ \label{eqn:f_S-max}
    }
  \end{align}
\end{definition}

We assume that the item and user vectors are derived from a Non-negative Matrix Factorization (NMF) model, ensuring a non-negative inner product $\ip{\bm{p},\bm{q}}$ for any $\bm{p}$ and $\bm{q}$.
This is crucial for establishing approximation bounds for \textsc{Greedy} and \textsc{DualGreedy}, as detailed in Section \ref{sect:methods}.
In Eqs.~\ref{eqn:f_S-avg} and~\ref{eqn:f_S-max}, we define $\frac{\lambda}{k} \sum_{\bm{p}\in \mathcal{S}} \ip{\bm{p}, \bm{q}}$ as the ``relevance term'' and $\frac{2\mu(1-\lambda)}{k(k-1)}$ $\sum_{\bm{p} \neq \bm{p}^\prime \in \mathcal{S}} \ip{\bm{p}, \bm{p}^\prime}$ (or $\mu(1-\lambda) \max_{\bm{p} \neq \bm{p}^\prime \in \mathcal{S}} \ip{\bm{p}, \bm{p}^\prime}$) as the ``diversity term.''
$k$MIPS is a special case of D$k$MIPS when $\lambda=1$, and D$k$MIPS can be transformed into the max-mean \cite{kuo1993analyzing} or max-min \cite{ravi1994heuristic} dispersion problem when $\lambda=0$.
Adjusting $\lambda$ between $0$ and $1$ allows for tuning the balance between relevance and diversity, with a smaller $\lambda$ favoring diversity and vice versa. 
According to Definition \ref{def:diverse-kmips}, D$k$MIPS employs the concept of MMR \cite{carbonell1998use}, which is known as an NP-hard problem \cite{drosou2010search}. Similarly, the max-mean and max-min dispersion problems are also NP-hard \cite{kuo1993analyzing, ravi1994heuristic}.
Therefore, we aim to develop approximation algorithms to find a result set $\mathcal{S}$ with a large $f(\mathcal{S})$ value.
Before presenting our algorithms, we summarize the frequently used notations in Table \ref{tab:notations}.

\section{Linear Scan-based Methods}
\label{sect:methods}

\subsection{\textsc{Greedy}}
\label{sect:methods:greedy}

Due to the fact that D$k$MIPS is grounded in the concept of MMR and drawing inspiration from established solutions for MMR-based problems \cite{kuo1993analyzing, VieiraRBHSTT11, AshkanKBW15},
we develop \textsc{Greedy}, a new greedy algorithm customized for the D$k$MIPS problem.

\begin{algorithm}[t]
\small
\caption{\textsc{Greedy}}
\label{alg:greedy}
\KwIn{Item set $\mathcal{P}$, query vector $\bm{q}$, integer $k \in \mathbb{Z}^{+}$, balancing factor $\lambda \in [0,1]$, scaling factor $\mu > 0$}
\KwOut{Set $\mathcal{S}$ of $k$ item vectors}
$\bm{p}^* \leftarrow \argmax_{\bm{p} \in \mathcal{P}} \ip{\bm{p}, \bm{q}}$\; \label{greedy:find-p-with-mip}
$\mathcal{S} \leftarrow \{\bm{p}^*\}$\; \label{greedy:add-p-with-mip}
\For{$i = 2$ \KwTo $k$}{ \label{greedy:stop-k}
  $\bm{p}^* \gets \argmax_{\bm{p} \in \mathcal{P} \setminus \mathcal{S}} \Delta_f(\bm{p}, \mathcal{S})$\; \label{greedy:find-p-with-max-marginal}
  $\mathcal{S} \leftarrow \mathcal{S} \cup \{\bm{p}^*\}$\; \label{greedy:add-p-with-max-marginal}
}
\Return $\mathcal{S}$\; \label{greedy:end}
\end{algorithm}

\paragraph{Algorithm Description}
The basic idea of \textsc{Greedy} is to iteratively find the item vector $\bm{p}^*$ that maximally increases the objective function $f(\cdot)$ w.r.t.~the current result set $\mathcal{S}$ and add it to $\mathcal{S}$.
We present its overall procedure in Algorithm \ref{alg:greedy}.
Specifically, at the first iteration, we add the item vector $\bm{p}^*$ that has the largest inner product with the query vector $\bm{q}$ to $\mathcal{S}$, i.e., $\bm{p}^* = \argmax_{\bm{p} \in \mathcal{P}} \ip{\bm{p},\bm{q}}$ (Lines \ref{greedy:find-p-with-mip} and \ref{greedy:add-p-with-mip}).
This is because the initial result set $\mathcal{S}$ is $\varnothing$, for which the diversity term is always $0$.
In subsequent iterations, we find the item vector $\bm{p}^*$ with the highest marginal gain $\Delta_{f}(\bm{p},\mathcal{S})$ (Line~\ref{greedy:find-p-with-max-marginal}), i.e., $\bm{p}^* = \argmax_{\bm{p} \in \mathcal{P}\setminus\mathcal{S}} \Delta_{f}(\bm{p},\mathcal{S})$, 
where $\Delta_f(\bm{p},\mathcal{S})$ for each $\bm{p} \in \mathcal{P} \setminus \mathcal{S}$ and $\mathcal{S}$ is computed from Eq.~\ref{eqn:incremental-f_S-avg} or~\ref{eqn:incremental-f_S-max}, each is derived from one of the two diversity measures:
\begin{align}
  & \Delta_{f_{avg}}(\bm{p},\mathcal{S}) = \tfrac{\lambda}{k} \ip{\bm{p},\bm{q}} - \tfrac{2\mu(1-\lambda)}{k(k-1)} \textstyle \sum_{\bm{p}' \in \mathcal{S}} \ip{\bm{p},\bm{p}'}, \label{eqn:incremental-f_S-avg} \\
  & \Delta_{f_{max}}(\bm{p},\mathcal{S}) = \tfrac{\lambda}{k} \ip{\bm{p},\bm{q}} - \mu(1-\lambda) \Delta_{max}(\bm{p}, \mathcal{S}), \label{eqn:incremental-f_S-max}
\end{align}
where $\Delta_{max}(\bm{p}, \mathcal{S}) = \max_{\bm{p}_x,\bm{p}_y \in \mathcal{S} \cup \{\bm{p}\} \land \bm{p}_x \neq \bm{p}_y} \ip{\bm{p}_x, \bm{p}_y} - \max_{\bm{p}_x, \bm{p}_y \in \mathcal{S} \land \bm{p}_x \neq \bm{p}_y} \ip{\bm{p}_x, \bm{p}_y}$.
Then, we add $\bm{p}^*$ to $\mathcal{S}$ (Line \ref{greedy:add-p-with-max-marginal}).
The iterative process ends when $\num{\mathcal{S}} = k$ (Line \ref{greedy:stop-k}), and $\mathcal{S}$ is returned as the D$k$MIPS result of the query $\bm{q}$ (Line \ref{greedy:end}).

The time complexity of \textsc{Greedy} is $O(ndk^2)$, regardless of whether it uses $f_{avg}(\cdot)$ or $f_{max}(\cdot)$.
It evaluates a maximum of $n$ items in each of the $k$ iterations.
For each item $\bm{p}$ in an iteration, it takes $O(d)$ time to compute $\ip{\bm{p}, \bm{q}}$, followed by an extra $O(kd)$ time to compute the diversity value.

\paragraph{Theoretical Analysis}
Next, using the sandwich strategy, we provide a \emph{data-dependent} approximation factor of \textsc{Greedy} for D$k$MIPS with both objective functions.
\vspace{-0.25em}
\begin{theorem}\label{theorem:approx-max}
  Let $\mathcal{S}$ be the D$k$MIPS result by \textnormal{\textsc{Greedy}} and $\mathcal{S}'$ be the $k$MIPS result for $\bm{q}$.
  Define $\overline{f}(\mathcal{S}) := \frac{\lambda}{k} \sum_{\bm{p} \in \mathcal{S}} \ip{\bm{p}, \bm{q}}$ and $\underline{f}(\mathcal{S}) := \frac{\lambda}{k} \sum_{\bm{p}\in \mathcal{S}} \ip{\bm{p}, \bm{q}} - div^*$, where $div^*= \max_{\mathcal{T} \subseteq \mathcal{P}, \num{\mathcal{T}} = k} \sum_{\bm{p}_x \neq \bm{p}_y \in \mathcal{T}}\tfrac{2\mu(1-\lambda)}{k(k-1)} \ip{\bm{p}_x, \bm{p}_y}$ for $f_{avg}(\cdot)$ or $\max_{\bm{p}_x \neq \bm{p}_y \in \mathcal{P}} \mu(1-\lambda) \ip{\bm{p}_x, \bm{p}_y}$ for $f_{max}(\cdot)$.
  Suppose that $f(\mathcal{S}') > 0$, $f(\mathcal{S}) \geq \max$ $(\tfrac{f(\mathcal{S}')}{\overline{f}(\mathcal{S}')} \cdot f(\mathcal{S}^*), f(\mathcal{S}^*) - div^*) - \Delta'$, where $\Delta' = \max(0, f(\mathcal{S}') - f(\mathcal{S}))$.
\end{theorem}
\vspace{-0.25em}
\begin{proof}
  We first provide the upper- and lower-bound functions of $f(\cdot)$.
  The upper bound function $\overline{f}(\cdot)$ is obtained simply by removing the diversity term from $f(\cdot)$ because the diversity term is always non-negative for any subset of $\mathcal{P}$.
  The lower bound function $\underline{f}$ is acquired by replacing the diversity term with its upper bound $div^*$, and $div^*$ is the maximum of the diversity function without considering the relevance term.
  The optimal results to maximize $\overline{f}$ and $\underline{f}$ are both the $k$MIPS results $\mathcal{S}'$.
  Thus, we have
  \begin{equation*}
    f(\mathcal{S}') = \tfrac{f(\mathcal{S}')}{\overline{f}(\mathcal{S}')} \cdot \overline{f}(\mathcal{S}') \geq \tfrac{f(\mathcal{S}')}{\overline{f}(\mathcal{S}')} \cdot \overline{f}(\mathcal{S}^*) \geq \tfrac{f(\mathcal{S}')}{\overline{f}(\mathcal{S}')} \cdot f(\mathcal{S}^*).
  \end{equation*}
  Recall that $\mathcal{S}^*$ is the optimal D$k$MIPS result.
  We have $f(\mathcal{S}') \geq \underline{f}(\mathcal{S}') \geq \underline{f}(\mathcal{S}^*) \geq f(\mathcal{S}^*) - div^*$.
  It is obvious that $f(\mathcal{S}) \geq f(\mathcal{S}') - \Delta'$, where $\Delta' = \max(0, f(\mathcal{S}') - f(\mathcal{S}))$.
  By combining all these results, Theorem~\ref{theorem:approx-max} is established.
\end{proof}

\subsection{\textsc{DualGreedy}}
\label{sect:methods:dual-greedy}

\noindent\textbf{Motivation.}
The approximation bound presented in Theorem \ref{theorem:approx-max} has several limitations.
It depends on the gap between the upper and lower bound functions, is valid only when $\lambda$ is close to $1$, and presumes the non-negativity of $f(\mathcal{S}')$.
To remedy these issues, we aim to establish a tighter bound based on the properties of the objective functions.

It is evident that $f_{avg}(\cdot)$ and $f_{max}(\cdot)$ are non-monotone because adding a highly similar item to existing ones in $\mathcal{S}$ can lead to a greater reduction in the diversity term compared to its contribution to the relevance term.
However, despite its non-monotonic nature, we show that $f_{avg}(\cdot)$ exhibits the celebrated property of \emph{submodularity} -- often referred to as the ``diminishing returns'' property -- where adding an item to a larger set yields no higher marginal gain than adding it to a smaller set \cite{KrauseG14}.
Formally, a set function $g: 2^{\mathcal{P}} \rightarrow \mathbb{R}$ on a ground set $\mathcal{P}$ is submodular if it fulfills $g(\mathcal{S} \cup \{\bm{p}\}) - g(\mathcal{S}) \geq g(\mathcal{T} \cup \{\bm{p}\}) - g(\mathcal{T})$ for any $\mathcal{S} \subseteq \mathcal{T} \subseteq \mathcal{P}$ and $\bm{p} \in \mathcal{P} \setminus \mathcal{T}$. 
In the following, we provide a formal proof of the submodularity of $f_{avg}(\cdot)$.
\vspace{-0.25em}
\begin{lemma}\label{lemma:avg-submodular}
  $f_{avg}: 2^{\mathcal{P}} \rightarrow \mathbb{R}$ is a submodular function.
\end{lemma}
\vspace{-0.25em}
\begin{proof}
  Let us define the marginal gain of adding an item vector $\bm{p}$ to a set $\mathcal{S}$ as $\Delta_{f_{avg}}(\bm{p},\mathcal{S}):= f_{avg}(\mathcal{S} \cup \{\bm{p}\}) - f_{avg}(\mathcal{S})$. The calculation of $\Delta_{f_{avg}}(\bm{p},\mathcal{S})$ is given in Eq.~\ref{eqn:incremental-f_S-avg}.
  When $\mathcal{S} = \varnothing$, $\Delta_{f_{avg}}(\bm{p}, \varnothing) = \frac{\lambda}{k} \ip{\bm{p},\bm{q}}$.
  When $\mathcal{S} \neq \varnothing$, for any $\bm{p} \in \mathcal{P} \setminus \mathcal{S}$, we have $\Delta_{f_{avg}}(\bm{p}, \mathcal{S}) = \frac{\lambda}{k} \ip{\bm{p},\bm{q}} - \frac{2\mu(1-\lambda)}{k(k-1)} \sum_{\bm{p}' \in \mathcal{S}} \ip{\bm{p},\bm{p}'} \leq \frac{\lambda}{k} \ip{\bm{p},\bm{q}} = \Delta_{f_{avg}}(\bm{p}, \varnothing)$ since the diversity term is non-negative.
  Furthermore, for any non-empty sets $\mathcal{S} \subseteq \mathcal{T} \subseteq \mathcal{P}$ and item $\bm{p} \in \mathcal{P} \setminus \mathcal{T}$,
  $\Delta_{f_{avg}}(\bm{p}, \mathcal{S}) - \Delta_{f_{avg}}(\bm{p}, \mathcal{T}) = \tfrac{2\mu(1-\lambda)}{k(k-1)} \textstyle \sum_{\bm{p}' \in \mathcal{T} \setminus \mathcal{S}} \ip{\bm{p},\bm{p}'} \geq 0$.
  In both cases, we confirm the submodularity of $f_{avg}$.
\end{proof}

Moreover, we call a set function $g: 2^{\mathcal{P}} \rightarrow \mathbb{R}$ supermodular if $g(\mathcal{S} \cup \{\bm{p}\}) - g(\mathcal{S}) \leq g(\mathcal{T} \cup \{\bm{p}\}) - g(\mathcal{T})$ for any $\mathcal{S} \subseteq \mathcal{T} \subseteq \mathcal{P}$ and $\bm{p} \in \mathcal{P} \setminus \mathcal{T}$.
We then use a counterexample to show that, unlike $f_{avg}(\cdot)$, $f_{max}(\cdot)$ is neither submodular nor supermodular.
\vspace{-0.25em}
\begin{example}\label{example:max-non-submoduler-and-supermodular}
  Consider a set $\mathcal{P}$ comprising four item vectors: $\bm{p}_1 = (1, 1)$, $\bm{p}_2 = (1, 0)$, $\bm{p}_3 = (2, 0)$, and $\bm{p}_4 = (0, 2)$, as well as a query $\bm{q} = (0.5, 0.5)$. 
  We set $\lambda = 0.5$, $k = 3$, and $\mu = \frac{1}{3}$.
  When examining $f_{max}: 2^{\mathcal{P}} \rightarrow \mathbb{R}$ on $\mathcal{P}$, we find that it is neither submodular nor supermodular.
  For example, considering two result sets $\mathcal{S}_{1} = \{\bm{p}_1\}$ and $\mathcal{S}_{12} = \{\bm{p}_1, \bm{p}_2\}$, $\Delta_{f_{max}}(\bm{p}_3, \mathcal{S}_{1}) = \frac{1}{6} (1 - 2) = -\frac{1}{6} < \Delta_{f_{max}}(\bm{p}_3, \mathcal{S}_{12}) = \frac{1}{6} (1 - 1) = 0$, thereby $f_{max}(\cdot)$ does not satisfy submodularity.
  On the other hand, when considering $\mathcal{S}_{2} = \{\bm{p}_2\}$ and $\mathcal{S}_{12} = \{\bm{p}_1, \bm{p}_2\}$, $\Delta_{f_{max}}(\bm{p}_4, \mathcal{S}_{1}) = \frac{1}{6} (1 - 0) = \frac{1}{6} > \Delta_{f_{max}}(\bm{p}_4, \mathcal{S}_{12}) = \frac{1}{6} (1 - 1) = 0$, indicating that $f_{max}(\cdot)$ is also not supermodular.
  \hfill $\triangle$ \par 
\end{example}

The results of Example~\ref{example:max-non-submoduler-and-supermodular} imply that maximizing $f_{max}(\cdot)$ from the perspective of submodular (or supermodular) optimization is infeasible.
Furthermore, \textsc{Greedy} cannot achieve a data-independent approximation factor to maximize $f_{avg}(\cdot)$ due to its non-monotonicity.
However, by exploiting the submodularity of $f_{avg}(\cdot)$, we propose \textsc{DualGreedy}, which enhances the greedy selection process by maintaining two result sets concurrently to effectively overcome the non-monotonicity barrier and achieve a \emph{data-independent} approximation for D$k$MIPS when $f_{avg}(\cdot)$ is used.

\paragraph{Algorithm Description}
The \textsc{DualGreedy} algorithm, depicted in Algorithm~\ref{alg:dual-greedy}, begins by initializing two result sets $\mathcal{S}_1$ and $\mathcal{S}_2$ as $\varnothing$ (Line \ref{dual-greedy:init}). 
At each iteration, we evaluate every item vector $\bm{p} \in \mathcal{P} \setminus (\mathcal{S}_1 \cup \mathcal{S}_2)$ that has not yet been added to $\mathcal{S}_1$ and $\mathcal{S}_2$.
For each $\mathcal{S}_i$ ($i \in \{1,2\}$), like \textsc{Greedy}, we find the item vector $\bm{p}_i^*$ with the largest marginal gain $\Delta_{f}(\bm{p}_i^*,\mathcal{S}_i)$ if the size of $\mathcal{S}_i$ has not reached $k$ (Lines \ref{dual-greedy:find-p:start}--\ref{dual-greedy:find-p:end}); 
Then, it adds $\bm{p}_1^*$ to $\mathcal{S}_1$ if $\Delta_f(\bm{p}^*_1, \mathcal{S}_1) \geq \Delta_f(\bm{p}^*_2, \mathcal{S}_2)$ or $\bm{p}_2^*$ to $\mathcal{S}_2$ otherwise (Line \ref{dual-greedy:add-p:start}--\ref{dual-greedy:add-p:end}).
The iterative process ends when the sizes of $\mathcal{S}_1$ and $\mathcal{S}_2$ reach $k$ (Line \ref{dual-greedy:stop-2k}) or $\Delta_f(\bm{p}^*_1, \mathcal{S}_1)$ and $ \Delta_f(\bm{p}^*_2, \mathcal{S}_2)$ are negative (Line \ref{dual-greedy:stop-early}). 
Finally, the one between $\mathcal{S}_1$ and $\mathcal{S}_2$ with the larger value of the objection function is returned as the D$k$MIPS answer for the query $\bm{q}$ (Line \ref{dual-greedy:end}).

\begin{algorithm}[t]
\small
\caption{\textsc{DualGreedy}}
\label{alg:dual-greedy}
\KwIn{Item set $\mathcal{P}$, query vector $\bm{q}$, integer $k \in \mathbb{Z}^{+}$, balancing factor $\lambda \in [0,1]$, scaling factor $\mu > 0$}
\KwOut{Set $\mathcal{S}$ of at most $k$ item vectors}
$\mathcal{S}_1, \mathcal{S}_2 \gets \varnothing$\; \label{dual-greedy:init}
\While{$|\mathcal{S}_1| < k$ or $|\mathcal{S}_2| < k$}{ \label{dual-greedy:stop-2k}
  $\bm{p}^*_1, \bm{p}^*_2 \gets NULL$\; \label{dual-greedy:find-p:start}
  \lIf{$|\mathcal{S}_1| < k$}{$\bm{p}^*_1 \gets \argmax_{\bm{p} \in \mathcal{P} \setminus (\mathcal{S}_1 \cup \mathcal{S}_2)} \Delta_f(\bm{p}, \mathcal{S}_1)$}
  \lIf{$|\mathcal{S}_2| < k$}{$\bm{p}^*_2 \gets \argmax_{\bm{p} \in \mathcal{P} \setminus (\mathcal{S}_1 \cup \mathcal{S}_2)} \Delta_f(\bm{p}, \mathcal{S}_2)$} \label{dual-greedy:find-p:end}
  \lIf{$\max_{j \in \{1, 2\}}\Delta_f(\bm{p}^*_j, \mathcal{S}_j) \leq 0$}{\textbf{break}} \label{dual-greedy:stop-early}
  \uIf{$\Delta_f(\bm{p}^*_1, \mathcal{S}_1)\geq \Delta_f(\bm{p}^*_2, \mathcal{S}_2)$}{ \label{dual-greedy:add-p:start}
    $\mathcal{S}_1 \leftarrow \mathcal{S}_1 \cup \{\bm{p}^*_1\}$\;
  }
  \Else{
    $\mathcal{S}_2 \leftarrow \mathcal{S}_2 \cup \{\bm{p}^*_2\}$\;
  } \label{dual-greedy:add-p:end}
}
\Return $\mathcal{S} \gets \argmax_{j \in \{1, 2\}} f(\mathcal{S}_j)$\; \label{dual-greedy:end}
\end{algorithm}

Like \textsc{Greedy}, \textsc{DualGreedy} takes $O(ndk^2)$ time for both $f_{avg}(\cdot)$ and $f_{max}(\cdot)$. 
It evaluates up to $2n$ items per iteration for $\mathcal{S}_1$ and $\mathcal{S}_2$, spans up to $2k$ iterations, and requires $O(kd)$ time per item to determine $\bm{p}_i^*$ for each $i \in \{1, 2\}$.

\paragraph{Theoretical Analysis}
Next, we analyze the approximation factor of \textsc{DualGreedy} for D$k$MIPS when $f_{avg}(\cdot)$ is used.
We keep two parallel result sets in \textsc{DualGreedy} to achieve a constant approximation factor for non-monotone submodular maximization.
However, the approximation factor is specific for \emph{non-negative} functions, whereas $f_{avg}(\cdot)$ is not.
In Theorem \ref{theorem:approx-avg}, we remedy the non-negativity of $f_{avg}(\cdot)$ by adding a regularization term and show that \textsc{DualGreedy} obtains an approximation ratio of $\frac{1}{4}$ after regularization.
\vspace{-0.25em}
\begin{theorem}\label{theorem:approx-avg}
  Let $\mathcal{S}$ be the D$k$MIPS result returned by \textnormal{\textsc{DualGreedy}}.
  We have $f_{avg}(\mathcal{S}) \geq \frac{1}{4} f_{avg}(\mathcal{S}^*) - \frac{3}{4} div^*_{max}$, where $div^*_{max} = \max_{\bm{p}_x, \bm{p}_y \in \mathcal{P} \wedge \bm{p}_x \neq \bm{p}_y}$ $\mu(1-\lambda) \ip{\bm{p}_x, \bm{p}_y}$.
\end{theorem}
\vspace{-0.25em}
\begin{proof}
  Define a function $f'_{avg}(\mathcal{T}) := f_{avg}(\mathcal{T}) + div^*_{max}$, where $div^*_{max} = \max_{\bm{p}_x, \bm{p}_y \in \mathcal{P} \wedge \bm{p}_x \neq \bm{p}_y} \mu(1-\lambda) \ip{\bm{p}_x, \bm{p}_y}$.
  The average pairwise inner product of any subset of $\mathcal{P}$ is bounded by $div^*_{max}$, ensuring $f'_{avg}(\mathcal{T}) \geq 0$ for any $\mathcal{T} \subseteq \mathcal{P}$.
  As the difference between $f'_{avg}(\cdot)$ and $f_{avg}(\cdot)$ is constant for a specific item set, $f'_{avg}(\cdot)$ remains a submodular function.
  Thus, \textsc{DualGreedy} yields identical results for both $f'_{avg}(\cdot)$ and $f_{avg}(\cdot)$ as the marginal gains remain constant for any item and subset.
  Moreover, the optimal solution for maximizing $f'_{avg}(\cdot)$ coincides with that for maximizing $f_{avg}(\cdot)$.
  As $f'_{avg}(\cdot)$ is non-negative, non-monotone, and submodular, according to \cite[Theorem~1]{HanCCW20}, we have $f'_{avg}(\mathcal{S}) \geq \frac{1}{4} f'_{avg}(\mathcal{S}^*)$.
  Since $f'_{avg}(\mathcal{S}) = f_{avg}(\mathcal{S}) + div^*_{max}$ and $f'_{avg}(\mathcal{S}^*) = f_{avg}(\mathcal{S}^*) + div^*_{max}$, we conclude the proof by subtracting $div^*_{max}$ from both sides.
\end{proof}

Theorem~\ref{theorem:approx-avg} is not applicable to $f_{max}(\cdot)$ due to its non-submodularity.
Yet, \textsc{DualGreedy} is not limited to $f_{avg}(\cdot)$ alone. By extending Theorem~\ref{theorem:approx-max}, it also attains data-dependent approximations for both $f_{avg}(\cdot)$ and $f_{max}(\cdot)$ and demonstrates strong performance. This will be validated with experimental results in Section \ref{sect:expt:recommendation}. 
The detailed extension methodologies are omitted due to space limitations. 


\subsection{Optimization}
\label{sect:methods:optimizations}

To improve efficiency, especially when $k$ is large, we introduce optimization techniques for \textsc{Greedy} and \textsc{DualGreedy} to prune unnecessary re-evaluations at every iteration.

\paragraph{Optimization for $f_{avg}(\cdot)$}
In each iteration $i$ ($2 \leq i \leq k$), the bottleneck in \textsc{Greedy} and \textsc{DualGreedy} is the need to compute $\Delta_{f_{avg}}(\bm{p},\mathcal{S})$ for each $\bm{p} \in \mathcal{P} \setminus \mathcal{S}$, which requires $O(kd)$ time.
Let $\bm{p}^j$ be the item added to $\mathcal{S}$ in the $j$-th iteration for some $j < i$.
We observe that $\sum_{\bm{p}' \in \mathcal{S}} \ip{\bm{p},\bm{p}'} = \textstyle \sum_{j=1}^{i-1} \ip{\bm{p},\bm{p}^j}$.
Thus, we maintain $div_{avg}(\bm{p}, \mathcal{S}) = \sum_{\bm{p}' \in \mathcal{S}} \ip{\bm{p},\bm{p}'}$ for each $\bm{p} \in \mathcal{P} \setminus \mathcal{S}$.
When adding $\bm{p}^j$ to $\mathcal{S}$, we can incrementally update $div_{avg}(\bm{p}, \mathcal{S})$ by adding $\ip{\bm{p},\bm{p}^j}$ in $O(d)$ time.
This enables the computation of marginal gains for all items in $O(nd)$ time per iteration, reducing the time complexity of \textsc{Greedy} and \textsc{DualGreedy} to $O(ndk)$.

\paragraph{Optimization for $f_{max}(\cdot)$}
Similarly to the computation of $\Delta_{f_{avg}}(\bm{p},\mathcal{S})$, computing $\Delta_{f_{max}}(\bm{p},\mathcal{S})$ needs $O(kd)$ time for each $\bm{p} \in \mathcal{P} \setminus \mathcal{S}$.
We find that $\max_{\bm{p}_x \neq \bm{p}_y \in \mathcal{S} \cup \{\bm{p}\}} \ip{\bm{p}_x, \bm{p}_y} = \max\{\max_{\bm{p}_x \in \mathcal{S}} \ip{\bm{p}, \bm{p}_x}, \max_{\bm{p}_x \neq \bm{p}_y \in \mathcal{S}} \ip{\bm{p}_x, \bm{p}_y}\}$, and it is evident that $\max_{\bm{p}_x \in \mathcal{S}} \ip{\bm{p}, \bm{p}_x} = \max\{ \ip{\bm{p}, \bm{p}^1}, \cdots, \ip{\bm{p}, \bm{p}^{i-1}}\}$.
Thus, we keep $div_{max}(\bm{p}, \mathcal{S}) = \max_{\bm{p}_x \in \mathcal{S}} \ip{\bm{p}, \bm{p}_x}$ for each $\bm{p} \in \mathcal{P} \setminus \mathcal{S}$. 
When adding $\bm{p}^j$ to $\mathcal{S}$, we update $div_{max}(\bm{p}, \mathcal{S})$ incrementally by comparing it with $\ip{\bm{p}, \bm{p}^j}$ in $O(d)$ time.
Since updating $\max_{\bm{p}_x \neq \bm{p}_y \in \mathcal{S}} \ip{\bm{p}_x, \bm{p}_y}$ only takes $O(1)$ time, \textsc{Greedy} and \textsc{DualGreedy} can be done in $O(ndk)$ time.

\section{Tree-based Methods}
\label{sect:bctree}

While \textsc{Greedy} and \textsc{DualGreedy} exhibit linear time complexities relative to $n$ and $k$, improving their efficiency is essential for seamless real-time user interactions.
The main challenge lies in scanning \emph{all} items in each iteration to find $\bm{p}^*$ with the largest marginal gain, i.e., $\bm{p}^* = \argmax_{\bm{p} \in \mathcal{P}\setminus\mathcal{S}} \Delta_{f}(\bm{p},\mathcal{S})$.
To address this issue, we integrate the BC-Tree index \cite{huang2023lightweight} into both algorithms.
This speeds up the identification of $\bm{p}^*$ by establishing a series of upper bounds for $\Delta_{f}(\bm{p}, \mathcal{S})$, resulting in \textsc{BC-Greedy} and \textsc{BC-DualGreedy}.

\subsection{Background: BC-Tree}
\label{sect:bctree:structure}

\noindent\textbf{BC-Tree Structure.}
We first review the structure of BC-Tree \cite{huang2023lightweight}, a variant of Ball-Tree \cite{omohundro1989five, samet2006foundations}.
In the BC-Tree, each node $\mathcal{N}$ contains a subset of item vectors, i.e., $\mathcal{N} \subseteq \mathcal{P}$. Specifically, $\mathcal{N} = \mathcal{P}$ if $\mathcal{N}$ is the root node. 
We denote $\num{\mathcal{N}}$ as the number of item vectors in $\mathcal{N}$.
Every node $\mathcal{N}$ within the BC-Tree has two children: $\mathcal{N}.\mathcal{L}$ and $\mathcal{N}.\mathcal{R}$.
It follows that $\num{\mathcal{N}.\mathcal{L}} + \num{\mathcal{N}.\mathcal{R}} = \num{\mathcal{N}}$ and $\mathcal{N}.\mathcal{L} \cap \mathcal{N}.\mathcal{R} = \varnothing$.

\paragraph{BC-Tree Construction}
The pseudocode of BC-Tree construction is outlined in Algorithm \ref{alg:bc_construct}. 
It takes a maximum leaf size $N_0$ and the item set $\mathcal{P}$ as input.
BC-Tree, like Ball-Tree, maintains ball structures (i.e., a ball center $\mathcal{N}.\bm{c}$ and a radius $\mathcal{N}.r$) for its internal and leaf nodes (Line \ref{bc:node:center-and-radius}).
Additionally, BC-Tree keeps ball and cone structures for each item vector $\bm{p}$ in its leaf node, facilitating point-level pruning.
For the ball structure, since all $\bm{p} \in \mathcal{N}$ share the same center $\mathcal{N}.\bm{c}$, it maintains their radii $\{r_{\bm{p}}\}_{\bm{p} \in \mathcal{N}}$ (Line \ref{bc:leaf:ball}). 
For the cone structure, BC-Tree stores the $l_2$-norm $\norm{\bm{p}}$ and the angle $\varphi_{\bm{p}}$ between each $\bm{p} \in \mathcal{N}$ and $\mathcal{N}.\bm{c}$. 
To perform D$k$MIPS efficiently, it computes and stores $\norm{\bm{p}}\cos \varphi_{\bm{p}}$ and $\norm{\bm{p}}\sin \varphi_{\bm{p}}$ (Lines \ref{bc:leaf:cone:start}--\ref{bc:leaf:cone:end}) and sorts all $\bm{p} \in \mathcal{N}$ in descending order of $r_{\bm{p}}$ (Line \ref{bc:leaf:sort_r_x}).
It adopts the seed-growth rule (Lines \ref{bc:split:start}--\ref{bc:split:end}) to split an internal node $\mathcal{N}$ with the furthest pivot pair $\bm{p}_l,\bm{p}_r \in \mathcal{N}$ (Line \ref{bc:internal:split}). 
Each $\bm{p} \in \mathcal{N}$ is assigned to its closer pivot, creating two subsets $\mathcal{N}_l$ and $\mathcal{N}_r$ (Line \ref{bc:internal:selection}). 
The BC-Tree is built recursively (Lines \ref{bc:internal:left-child}--\ref{bc:internal:right-child}) until all leaf nodes contain at most $N_0$ items.

Below, Theorem \ref{theorem:ball_tree_construction} shows that the BC-Tree construction is fast ($\tilde{O}(dn)$ time) and lightweight ($O(nd)$ space) \cite{huang2023lightweight}.
\vspace{-0.25em}
\begin{theorem}\label{theorem:ball_tree_construction}
  The BC-Tree is constructed in $O(nd \log n)$ time and stored in $O(nd)$ space.
\end{theorem}

\subsection{Upper Bounds for Marginal Gains}
\label{sect:bctree:bounds}

\noindent\textbf{Node-Level Pruning.}
To find $\bm{p}^* = \argmax_{\bm{p} \in \mathcal{P}\setminus\mathcal{S}} \Delta_{f}(\bm{p},\mathcal{S})$ efficiently, we first establish an upper bound for the marginal gain $\Delta_{f}(\bm{p},\mathcal{S})$ using the ball structure.
\vspace{-0.25em}
\begin{theorem}[Node-Level Ball Bound]\label{theorem:node_upper_bound}
  Given a query vector $\bm{q}$ and an (internal or leaf) node $\mathcal{N}$ that maintains a center $\bm{c}$ and a radius $r$, the maximum possible $\Delta_f(\bm{p}, \mathcal{S})$ of all $\bm{p} \in \mathcal{N}$ and $\mathcal{S}$ defined by Eqs.~\ref{eqn:incremental-f_S-avg} and \ref{eqn:incremental-f_S-max} are bounded by Eq.~\ref{eqn:node_upper_bound-max}:
  \begin{equation}
    \textstyle \max_{\bm{p} \in \mathcal{N}} \Delta_{f}(\bm{p},\mathcal{S}) \leq \tfrac{\lambda}{k}(\ip{\bm{q}, \bm{c}} + r\norm{\bm{q}}). \label{eqn:node_upper_bound-max}
  \end{equation}
\end{theorem}
\vspace{-0.25em}
\begin{proof}
  We first consider $\Delta_{f_{avg}}(\bm{p},\mathcal{S})$.
  According to \cite[Theorem 3.1]{RamG12}, given the center $\bm{c}$ and radius $r$ of the ball, the maximum possible inner product between $\bm{p} \in \mathcal{N}$ and $\bm{q}$ is bounded by $\max_{\bm{p} \in \mathcal{N}} \ip{\bm{p},\bm{q}} \leq \ip{\bm{q}, \bm{c}} + r\norm{\bm{q}}$.
  Moreover, from Eq.~\ref{eqn:incremental-f_S-avg}, since the diversity term $\tfrac{2\mu(1-\lambda)}{k(k-1)} \sum_{\bm{p}' \in \mathcal{S}} \ip{\bm{p},\bm{p}'}$ is non-negative, $\Delta_{f_{avg}}(\bm{p},\mathcal{S}) \leq \frac{\lambda}{k} \ip{\bm{p}, \bm{q}}$.
  Therefore, $\max_{\bm{p} \in \mathcal{N}} \Delta_{f_{avg}}(\bm{p},\mathcal{S}) \leq \frac{\lambda}{k} \max_{\bm{p} \in \mathcal{N}} \ip{\bm{p}, \bm{q}} \leq \frac{\lambda}{k}(\ip{\bm{q}, \bm{c}} + r\norm{\bm{q}})$.
  A similar proof follows for $\Delta_{f_{max}}(\bm{p},\mathcal{S})$, and details are omitted for brevity.
\end{proof}

\begin{algorithm}[t]
\caption{\textsc{BCTreeConstruct}}
\label{alg:bc_construct}
\small
\KwIn{Item set $\mathcal{N} \subseteq \mathcal{P}$, maximum leaf size $N_0$}
\KwOut{(Internal or leaf) node $\mathcal{N}$}
$\mathcal{N}.\bm{c} \gets \tfrac{1}{\num{\mathcal{N}}} \sum_{\bm{p} \in \mathcal{N}} \bm{p}$;~$\mathcal{N}.r \gets \max_{\bm{p} \in \mathcal{N}} \norm{\bm{p}-\mathcal{N}.\bm{c}}$\; \label{bc:node:center-and-radius}
\eIf(\Comment*[f]{leaf node}){$\num{\mathcal{N}} \leq N_0$} {
  \ForEach{$\bm{p} \in \mathcal{N}$} {
    $r_{\bm{p}} \gets \norm{\bm{p}-\mathcal{N}.\bm{c}}$\; \label{bc:leaf:ball}
    $\norm{\bm{p}}\cos\varphi_{\bm{p}} \gets \ip{\bm{p}, \mathcal{N}.\bm{c}} / \norm{\mathcal{N}.\bm{c}}$\; \label{bc:leaf:cone:start}
    $\norm{\bm{p}}\sin\varphi_{\bm{p}} \gets \sqrt{\norm{\bm{p}}^2 - (\norm{\bm{p}}\cos\varphi_{\bm{p}})^2}$\; \label{bc:leaf:cone:end}
  }
  Sort all $\bm{p} \in \mathcal{N}$ in descending order of $r_{\bm{p}}$\;  \label{bc:leaf:sort_r_x}
}
(\Comment*[f]{internal node}){
  $\bm{p}_l, \bm{p}_r \leftarrow $~\texttt{Split}($\mathcal{N}$)\; \label{bc:internal:split}
  $\mathcal{N}_l \leftarrow \{\bm{p} \in \mathcal{N} \mid \norm{\bm{p}-\bm{p}_l} \leq \norm{\bm{p}-\bm{p}_r}\}$;~$\mathcal{N}_r \leftarrow \mathcal{N} \setminus \mathcal{N}_l$\; \label{bc:internal:selection}
  $\mathcal{N}.\mathcal{L} \leftarrow $~\textsc{BCTreeConstruct}($\mathcal{N}_l, N_0$)\; \label{bc:internal:left-child}
  $\mathcal{N}.\mathcal{R} \leftarrow $~\textsc{BCTreeConstruct}($\mathcal{N}_r, N_0$)\; \label{bc:internal:right-child}
}
\Return $\mathcal{N}$\;
\BlankLine
\SetKwFunction{Function}{Split}
\SetKwProg{Fn}{Function}{:}{}
\Fn{\Function{$\mathcal{N}$}}{
  Select a random item vector $\bm{v} \in \mathcal{N}$\; \label{bc:split:start}
  $\bm{p}_l \gets \argmax_{\bm{p} \in \mathcal{N}} \norm{\bm{p}-\bm{v}}$\;
  $\bm{p}_r \gets \argmax_{\bm{p} \in \mathcal{N}} \norm{\bm{p}-\bm{p}_l}$\;
  \Return $\bm{p}_l, \bm{p}_r$\; \label{bc:split:end}
}
\end{algorithm}

\paragraph{Point-Level Pruning}
To facilitate point-level pruning, it is intuitive to employ the ball structure for each item $\bm{p}$ in the leaf node $\mathcal{N}$. 
This allows us to obtain an upper bound of $\Delta_{f}(\bm{p},\mathcal{S})$ for each $\bm{p} \in \mathcal{N}$, as presented in Corollary \ref{corollary:point_upper_bound_ball}.
\vspace{-0.25em}
\begin{corollary}[Point-Level Ball Bound]\label{corollary:point_upper_bound_ball}
  Given a query vector $\bm{q}$ and a leaf node $\mathcal{N}$ that maintains a center $\bm{c}$ and the radius $r_{\bm{p}}$ for each item vector $\bm{p} \in \mathcal{N}$, the maximum possible $\Delta_f(\bm{p},\mathcal{S})$ defined by Eqs.~\ref{eqn:incremental-f_S-avg} and \ref{eqn:incremental-f_S-max} are bounded by Eq.~\ref{eqn:leaf_upper_bound-ball-max}:
  \begin{equation}
    \Delta_{f}(\bm{p},\mathcal{S}) \leq \tfrac{\lambda}{k}(\ip{\bm{q}, \bm{c}} + r_{\bm{p}} \norm{\bm{q}}). \label{eqn:leaf_upper_bound-ball-max}
  \end{equation}
\end{corollary}

Corollary \ref{corollary:point_upper_bound_ball} follows directly from Theorem \ref{theorem:node_upper_bound}. We omit the proof for brevity.
We call the Right-Hand Side (RHS) of Eq.~\ref{eqn:leaf_upper_bound-ball-max} as the \emph{point-level ball bound} of $\Delta_{f}(\bm{p},\mathcal{S})$. 
Since $\ip{\bm{q}, \bm{c}}$ has been computed when visiting $\mathcal{N}$, this bound can be computed in $O(1)$ time.
Moreover, as $\ip{\bm{q},\bm{c}}$ and $\norm{\bm{q}}$ are fixed for a given $\bm{q}$, it increases as $r_{\bm{p}}$ increases. 
Thus, when constructing $\mathcal{N}$, we sort all $\bm{p} \in \mathcal{N}$ in descending order of $r_{\bm{p}}$ (as depicted in Line \ref{bc:leaf:sort_r_x} of Algorithm \ref{alg:bc_construct}), thus pruning the items $\bm{p} \in \mathcal{N}$ \emph{in a batch}.
Yet, the point-level ball bounds might not be tight enough. Let $\theta$ be the angle between $\bm{c}$ and $\bm{q}$. Next, we utilize the cone structure (i.e., the $l_2$-norm $\norm{\bm{p}}$ and its angle $\varphi_{\bm{p}}$ for each $\bm{p} \in \mathcal{N}$) to obtain a tighter upper bound. 
\vspace{-0.25em}
\begin{theorem}[Point-Level Cone Bound]\label{theorem:point_lower_bound_cone}
  Given a query vector $\bm{q}$ and a leaf node $\mathcal{N}$ that maintain the $l_2$-norm $\norm{\bm{p}}$ and the angle $\varphi_{\bm{p}}$ for each item vector $\bm{p} \in \mathcal{N}$, the maximum possible $\Delta_f(\bm{p}, \mathcal{S})$ defined by Eqs.~\ref{eqn:incremental-f_S-avg} and \ref{eqn:incremental-f_S-max} are bounded by Eq.~\ref{eqn:leaf_upper_bound-cone-max}:
  \begin{equation}
    \Delta_{f}(\bm{p},\mathcal{S}) \leq \tfrac{\lambda}{k}(\norm{\bm{p}} \norm{\bm{q}} \cos(\abso{\theta - \varphi_{\bm{p}}})). \label{eqn:leaf_upper_bound-cone-max}
  \end{equation}
\end{theorem}
\vspace{-0.25em}
\begin{proof}
  Let us first consider $\Delta_{f_{avg}}(\bm{p},\mathcal{S})$. Suppose that $\theta_{\bm{p},\bm{q}}$ is the angle between $\bm{p}$ and $\bm{q}$. 
  According to the triangle inequality, $0 \leq \abso{\theta - \varphi_{\bm{p}}} \leq \theta_{\bm{p},\bm{q}} \leq \theta + \varphi_{\bm{p}} \leq 2\pi$. 
  Thus, $\ip{\bm{p},\bm{q}} = \norm{\bm{p}} \norm{\bm{q}} \cos\theta_{\bm{p},\bm{q}} \leq \norm{\bm{p}} \norm{\bm{q}} \max_{\theta_{\bm{p},\bm{q}} \in [\abso{\theta - \varphi_{\bm{p}}}, \theta + \varphi_{\bm{p}}]} \cos \theta_{\bm{p},\bm{q}}$.
  As $\cos\theta = \cos(2\pi - \theta)$, and for any $\theta \in [0,\pi]$, $\cos\theta$ decreases monotonically as $\theta$ increases, the upper bound of $\cos\theta_{\bm{p},\bm{q}}$ is either $\cos(\abso{\theta - \varphi_{\bm{p}}})$ or $\cos(\theta + \varphi_{\bm{p}})$. 
  On the other hand, as $0 \leq \theta,\varphi_{\bm{p}} \leq \pi$, $\sin \theta \geq 0$ and $\sin \varphi_{\bm{p}} \geq 0$. 
  Thus, $\cos(\theta+\varphi_{\bm{p}}) = \cos\theta \cos\varphi_{\bm{p}} - \sin\theta \sin\varphi_{\bm{p}} \leq \cos\theta \cos\varphi_{\bm{p}} + \sin\theta \sin\varphi_{\bm{p}}  = \cos(\abso{\theta-\varphi_{\bm{p}}})$, thereby $\ip{\bm{p},\bm{q}} \leq \norm{\bm{p}} \norm{\bm{q}} \cos(\abso{\theta - \varphi_{\bm{p}}})$.
  Based on Theorem \ref{theorem:node_upper_bound}, $\Delta_{f_{avg}}(\bm{p},\mathcal{S}) \leq \frac{\lambda}{k} \ip{\bm{p}, \bm{q}}$ for any $\bm{p} \in \mathcal{N}$. 
  Thus, $\Delta_{f_{avg}}(\bm{p},\mathcal{S}) \leq \frac{\lambda}{k} \norm{\bm{p}} \norm{\bm{q}} \cos(\abso{\theta - \varphi_{\bm{p}}})$.
  The above result also holds for $\Delta_{f_{max}}(\bm{p},\mathcal{S})$ by applying the same analytical procedure. We omit details for the sake of conciseness. 
\end{proof}

\begin{figure}[h]
\centering
\vspace{-0.75em}
\includegraphics[width=0.24\textwidth]{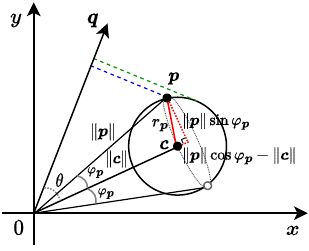}
\vspace{-0.75em}
\caption{Illustration of point-level ball bound (green dashed line) and point-level cone bound (blue dashed line) for an item vector $\bm{p}$ in the leaf node. From the red triangle, we observe that $(\norm{\bm{p}} \sin\varphi_{\bm{p}})^2 + (\norm{\bm{c}} - \norm{\bm{p}} \cos\varphi_{\bm{p}})^2 = r_{\bm{p}}^2$.}
\label{fig:point_ball_cone}
\vspace{-1.0em}
\end{figure}

We call the RHS of Eq.~\ref{eqn:leaf_upper_bound-cone-max} as the \emph{point-level cone bound} of $\Delta_{f}(\bm{p},\mathcal{S})$.
Note that $\norm{\bm{p}} \norm{\bm{q}} \cos(\abso{\theta-\varphi_{\bm{p}}}) = \norm{\bm{q}}\cos\theta \cdot \norm{\bm{p}}\cos\varphi_{\bm{p}} + \norm{\bm{q}}\sin\theta \cdot \norm{\bm{p}}\sin\varphi_{\bm{p}}$. 
To allow efficient pruning based on Eq.~\ref{eqn:leaf_upper_bound-cone-max}, we store $\norm{\bm{p}} \cos\varphi_{\bm{p}}$ and $\norm{\bm{p}} \sin\varphi_{\bm{p}}$ for each $\bm{p} \in \mathcal{N}$ when building the BC-Tree. 
Moreover, as $\ip{\bm{q},\bm{c}}$ has been computed when we visit the leaf node $\mathcal{N}$, $\norm{\bm{q}}\cos\theta = \ip{\bm{q},\bm{c}}/\norm{\bm{c}}$ and $\norm{\bm{q}}\sin\theta = \norm{\bm{q}} \sqrt{1 - \cos^2 \theta}$ can be computed in $O(1)$ time. Thus, the point-level cone bound can be computed in $O(1)$ time.
We show that the point-level cone bound is tighter than the point-level ball bound.
\vspace{-0.25em}
\begin{theorem}\label{theorem:tight_lower_bound}
  Given a query vector $\bm{q}$, for any item vector $\bm{p}$ in a leaf node $\mathcal{N}$, its point-level cone bound is strictly tighter than its point-level ball bound for $\Delta_{f}(\bm{p},\mathcal{S})$.
\end{theorem}
\vspace{-0.25em}
\begin{proof}
  To prove this, we need to show that the RHS of Eq.~\ref{eqn:leaf_upper_bound-ball-max} is at least the RHS of Eq.~\ref{eqn:leaf_upper_bound-cone-max}. 
  By removing $\tfrac{\lambda}{k}$ and $\norm{\bm{q}}$, which do not affect the inequality, we should show whether $\norm{\bm{c}}\cos\theta + r_{\bm{p}} \geq \norm{\bm{p}} \cos(\abso{\theta - \varphi_{\bm{p}}})$.
  Here is the calculation:
  \begin{displaymath}
    \resizebox{0.99\hsize}{!}{
    $\begin{aligned}
      & \norm{\bm{c}}\cos\theta + r_{\bm{p}} - \norm{\bm{p}} \cos(\abso{\theta - \varphi_{\bm{p}}}) \\
      & = r_{\bm{p}} - (\norm{\bm{p}}\sin\varphi_{\bm{p}} \cdot \sin\theta + (\norm{\bm{p}} \cos\varphi_{\bm{p}} - \norm{\bm{c}}) \cdot \cos\theta) \\
      & \geq r_{\bm{p}} - \textstyle \sqrt{(\norm{\bm{p}} \sin\varphi_{\bm{p}})^2 + (\norm{\bm{p}} \cos\varphi_{\bm{p}}-\norm{\bm{c}})^2} \cdot \textstyle \sqrt{\sin^2\theta + \cos^2\theta} \\ 
      & = r_{\bm{p}} - r_{\bm{p}} = 0,
    \end{aligned}$
    }
  \end{displaymath} 
  where the second last step is based on the Cauchy-Schwarz inequality and the last step relies on Pythagoras' theorem, which enables us to illustrate the relationship between the distances $\norm{\bm{p}} \sin\varphi_{\bm{p}}$, $\norm{\bm{p}} \cos\varphi_{\bm{p}}-\norm{\bm{c}}$, and $r_{\bm{p}}$. A visual depiction of this triangle (in red) is given in Fig.~\ref{fig:point_ball_cone}.
\end{proof}

\begin{algorithm}[t]
\small
\caption{\textsc{BCTreeSearch}}
\label{alg:bc_search}
\KwIn{Query vector $\bm{q} \in \mathbb{R}^d$, root node $\mathcal{N}_{root}$ of BC-Tree, integer $k \in \mathbb{Z}^{+}$, balancing factor $\lambda \in [0,1]$, scaling factor $\mu > 0$, and current result set $\mathcal{S}$}
\KwOut{Item $\bm{p}^*$ and the largest marginal gain $\tau$ w.r.t.~$\mathcal{S}$}
Initialize $\bm{p}^* \gets NULL$, $\tau \gets -\infty$, and $ip \gets \ip{\bm{q}, \mathcal{N}.\bm{c}}$\;\label{bc:main:init}
$\{\bm{p}^*,\tau\} \gets$~\texttt{SubtreeSearch}($\bm{p}^*, \tau, ip, \mathcal{N}_{root}$)\;\label{bc:main:branch}
\Return $\bm{p}^*$\; \label{bc:main:end}

\BlankLine
\SetKwFunction{Function}{SubtreeSearch}
\SetKwProg{Fn}{Function}{:}{}
\Fn{\Function{$\bm{p}^*, \tau, ip, \mathcal{N}$}}{ \label{bc:node:start}
  Compute ${ub}_{node}$ by the RHS of Eq.~\ref{eqn:node_upper_bound-max}\;\label{bc:node:ub}
  \lIf{$\tau \geq {ub}_{node}$}{\Return $\{\bm{p}^*,\tau\}$} \label{bc:node:early-stop}
  \eIf{$\num{\mathcal{N}} \leq N_0$}{
    $\{\bm{p}^*,\tau\} \gets$~\texttt{FilterScan}($\bm{p}^*, \tau, ip, \mathcal{N}$)\;\label{bc:node:scan}
  }{
    ${ip}_l \gets \ip{\bm{q}, \mathcal{N}.\mathcal{L}.\bm{c}}$;~${ip}_r \gets \tfrac{\num{\mathcal{N}}}{\num{\mathcal{N}.\mathcal{R}}} \cdot {ip} - \tfrac{\num{\mathcal{N}.\mathcal{L}}}{\num{\mathcal{N}.\mathcal{R}}} \cdot {ip}_l$\; \label{bc:node:branch-start}
    \eIf{${ip}_l \geq {ip}_r$} {
      $\{\bm{p}^*,\tau\} \gets$~\texttt{SubtreeSearch}($\bm{p}^*, \tau, {ip}_l, \mathcal{N}.\mathcal{L}$)\;
      $\{\bm{p}^*,\tau\} \gets$~\texttt{SubtreeSearch}($\bm{p}^*, \tau, {ip}_r, \mathcal{N}.\mathcal{R}$)\;
    }{
      $\{\bm{p}^*,\tau\} \gets$~\texttt{SubtreeSearch}($\bm{p}^*, \tau, {ip}_r, \mathcal{N}.\mathcal{R}$)\;
      $\{\bm{p}^*,\tau\} \gets$~\texttt{SubtreeSearch}($\bm{p}^*, \tau, {ip}_l, \mathcal{N}.\mathcal{L}$)\;
    } \label{bc:node:branch-end}
  }
  \Return $\{\bm{p}^*,\tau\}$\; \label{bc:node:end}
}

\BlankLine
\SetKwFunction{Function}{FilterScan} 
\SetKwProg{Fn}{Function}{:}{}
\Fn{\Function{$\bm{p}^*, \tau, ip, \mathcal{N}$}}{ \label{bc:scan:start}
  $\norm{\bm{q}}\cos\theta \gets ip/\norm{\mathcal{N}.\bm{c}}$;~$\norm{\bm{q}}\sin\theta \gets \norm{\bm{q}} \sqrt{1 - \cos^2 \theta}$\; \label{bc:scan:cone-para}
  \ForEach{$\bm{p} \in \mathcal{N}$}{
    Compute ${ub}_{ball}$ by the RHS of Eq.~\ref{eqn:leaf_upper_bound-ball-max}\; \label{bc:scan:ball-bounds}
    \lIf{$\tau \geq {ub}_{ball}$}{\textbf{break}} \label{bc:scan:early-stop}
    Compute ${ub}_{cone}$ by the RHS of Eq.~\ref{eqn:leaf_upper_bound-cone-max}\; \label{bc:scan:cone-bounds}
    \lIf{$\tau \geq {ub}_{cone}$}{\textbf{continue}} \label{bc:scan:skip}
    \If{$\frac{\lambda}{k}\ip{\bm{p},\bm{q}} < \tau$}{\label{bc:scan:calc-ip}
      Compute $\Delta_{f}(\bm{p},\mathcal{S})$ by Eq.~\ref{eqn:incremental-f_S-avg} or \ref{eqn:incremental-f_S-max}\; \label{bc:scan:calc-marginal-gain}
      \If{$\Delta_{f}(\bm{p},\mathcal{S}) > \tau$}{$\bm{p}^* \gets \bm{p}$ and $\tau \gets \Delta_{f}(\bm{p},\mathcal{S})$} \label{bc:scan:update}
    }
  }
  \Return $\{\bm{p}^*,\tau\}$\; \label{bc:scan:end}
}
\end{algorithm}

\subsection{\textsc{BC-Greedy} and \textsc{BC-DualGreedy}}
\label{sect:bctree:search}

We now introduce \textsc{BC-Greedy} and \textsc{BC-DualGreedy}, which follow procedures similar to \textsc{Greedy} and \textsc{DualGreedy}, respectively, and incorporate their optimization techniques.
The key distinction is the use of upper bounds derived from the BC-Tree structure, which allows for efficient identification of $\bm{p}^* = \argmax_{\bm{p} \in \mathcal{P} \setminus \mathcal{S}} \Delta_{f}(\bm{p},\mathcal{S})$.

\paragraph{Algorithm Description}
The BC-Tree search scheme to identify $\bm{p}^*$ is outlined in Algorithm \ref{alg:bc_search}. 
We initialize $\bm{p}^*$ as $NULL$ and the largest marginal gain $\tau$ as $-\infty$. 
Then, we call the procedure \texttt{SubtreeSearch} from the root node $\mathcal{N}_{root}$ to find $\bm{p}^*$ and update $\tau$, eventually returning $\bm{p}^*$ as a result.

In \texttt{SubtreeSearch}, we find $\bm{p}^*$ and update $\tau$ by traversing the BC-Tree in a depth-first manner.
For each node $\mathcal{N}$, we first compute its node-level ball bound ${ub}_{node}$ by the RHS of Eq.~\ref{eqn:node_upper_bound-max} (Line \ref{bc:node:ub}). 
If $\tau \geq {ub}_{node}$, we prune this branch as it cannot contain any item $\bm{p}$ having larger $\Delta_f(\bm{p}, \mathcal{S})$ than $\tau$ (Line \ref{bc:node:early-stop});
otherwise, we compute the inner products ${ip}_l$ and ${ip}_r$ between $\bm{q}$ and the centers from its left child $\mathcal{N}.\mathcal{L}$ and right child $\mathcal{N}.\mathcal{R}$, and recursively visit its two branches (Lines \ref{bc:node:branch-start}--\ref{bc:node:branch-end}). 
If $\mathcal{N}$ is a leaf node, we call the procedure \texttt{FilterScan} to update $\bm{p}^*$ and $\tau$ (Line \ref{bc:node:scan}).
After the traversal on the current node $\mathcal{N}$, $\bm{p}^*$ and $\tau$ are returned (Line \ref{bc:node:end}).

In \texttt{FilterScan}, we perform point-level pruning to efficiently identify $\bm{p}^*$. 
We start by computing $\norm{\bm{q}}\cos\theta$ and $\norm{\bm{q}}\sin\theta$ (Line \ref{bc:scan:cone-para}) so that the point-level cone bound $ub_{cone}$ can be computed in $O(1)$ time.
For each $\bm{p} \in \mathcal{N}$, we first compute the point-level ball bound ${ub}_{ball}$ using the RHS of Eq.~\ref{eqn:leaf_upper_bound-ball-max} (Line \ref{bc:scan:ball-bounds}). 
By scanning items in descending order of $r_{\bm{p}}$, we can prune $\bm{p}$ and the remaining items in a batch if $\tau \geq {ub}_{ball}$ (Line \ref{bc:scan:early-stop}).
If it fails, we compute ${ub}_{cone}$ by the RHS of Eq.~\ref{eqn:leaf_upper_bound-cone-max} (Line \ref{bc:scan:cone-bounds}) and skip the marginal gain $\Delta_f(\bm{p}, \mathcal{S})$ computation if $\tau \geq {ub}_{cone}$ (Line \ref{bc:scan:skip}).
If this bound also fails, we check whether $\frac{\lambda}{k}\ip{\bm{p},\bm{q}} < \tau$ (Line \ref{bc:scan:calc-ip}).
If yes, we take extra $O(d)$ time to compute $\Delta_f(\bm{p}, \mathcal{S})$ and update $\bm{p}^*$ and $\tau$ as $\bm{p}$ and $\Delta_f(\bm{p}, \mathcal{S})$, respectively, if $\Delta_f(\bm{p}, \mathcal{S}) > \tau$ (Lines \ref{bc:scan:calc-marginal-gain}--\ref{bc:scan:update}).
Finally, we return $\bm{p}^*$ and $\tau$ as answers (Line \ref{bc:scan:end}).

Like \textsc{Greedy} and \textsc{DualGreedy}, \textsc{BC-Greedy} and \textsc{BC-DualGreedy} process D$k$MIPS queries in $O(ndk)$ time and $O(n)$ space. 
This is because they inherit the same optimizations, and BC-Tree degrades to $O(nd)$ time in the worst case due to the necessity of examining all leaves and items.
Yet, in practice, the actual query time often falls below $O(ndk)$ and even $O(nd)$, owing to the effective node- and point-level pruning techniques that vastly enhance scalability. 
This impact will be substantiated in Sections \ref{sect:expt:recommendation}, \ref{sect:expt:query}, and \ref{sect:expt:scalability}.

\paragraph{Theoretical Analysis}
\textsc{BC-Greedy} and \textsc{BC-DualGreedy} always provide query results identical to \textsc{Greedy} and \textsc{DualGreedy}.
This is because they identify the same item $\bm{p}^*$ in each iteration.
Details are omitted for the sake of conciseness, but this will be validated in Section \ref{sect:expt:query}.



\section{Experiments}
\label{sect:expt}

\subsection{Experimental Setup}
\label{sect:expt:setup}

\noindent\textbf{Data Sets and Queries.}
We employ ten popular recommendation data sets for performance evaluation, including 
MovieLens,\footnote{\url{https://grouplens.org/datasets/movielens/25m/}} 
Netflix,\footnote{\url{https://www.kaggle.com/datasets/netflix-inc/netflix-prize-data}} 
Yelp,\footnote{\url{https://www.yelp.com/dataset}}
Food \cite{li2019food}, 
Google Local Data (2021) for California (Google-CA),\footnote{\url{https://datarepo.eng.ucsd.edu/mcauley_group/gdrive/googlelocal/}} 
Yahoo! Music C15 (Yahoo!),\footnote{\url{https://webscope.sandbox.yahoo.com/catalog.php?datatype=c}} 
Taobao,\footnote{\url{https://tianchi.aliyun.com/dataset/649}}
Amazon Review Data (2018) for the categories of `Books' (AMZ-Books) and `Clothing, Shoes, and Jewelry' (AMZ-CSJ),\footnote{\url{https://nijianmo.github.io/amazon/index.html}}
and LFM.\footnote{\url{http://www.cp.jku.at/datasets/LFM-1b/}}
For each data set with user ratings, we obtain its (sparse) rating matrix $\bm{R}$, where $\bm{R}(i,j)$ represents user $i$'s rating on item $j$, and normalize it to a 0-5 point scale.
For non-rating data sets, i.e., Taobao and LFM, we assume each user has a rating of 5 on all interacted items.
We set a latent dimension of $d =1 00$ and apply NMF \cite{fevotte2011algorithms, scikit-learn} on $\bm{R}$ to acquire the latent user and item matrices.
All near-zero user and item vectors are removed.
Finally, we randomly pick $100$ user vectors as the query set.

Moreover, we utilize the MovieLens, Yelp, Google-CA, and Yahoo! data sets, where the category (or genre) information is available, for recommendation quality evaluation.
The Taobao and Food data sets also have category information, but Taobao is excluded due to inconsistent item-to-category mapping, where one item can be linked to different categories across transactions, 
and Food is omitted because of its highly duplicated tags, which fail to accurately reflect recipe features.
The statistics of the ten data sets are presented in Table \ref{table:dataset}.

\paragraph{Benchmark Methods}
We compare the following methods for D$k$MIPS in the experiments.
\begin{itemize}[nolistsep]
  \item \textbf{\textsc{Linear} (LIN)} is a baseline that evaluates all items one by one to identify the $k$ items that have the largest inner products with the query vector in $O(nd)$ time.
  \item \textbf{\textsc{IP-Greedy} (IP-G)} \cite{hirata2022solving} is a D$k$MIPS method with a time complexity of $O(ndk^2\log n)$. We used its implementation published by the original authors.\footnote{\url{https://github.com/peitaw22/IP-Greedy}}
  \item \textbf{\textsc{Greedy} (G)} and \textbf{\textsc{DualGreedy} (D)} are the basic versions proposed in Sections \ref{sect:methods:greedy} and \ref{sect:methods:dual-greedy}, respectively, with the same time complexity of $O(ndk^2)$.
  \item \textbf{\textsc{Greedy}$^+$ (G$^+$)} and \textbf{\textsc{DualGreedy}$^+$ (D$^+$)} are improved versions of \textsc{Greedy} and \textsc{DualGreedy} by incorporating the optimization techniques outlined in Section \ref{sect:methods:optimizations} with a reduced time complexity of $O(ndk)$.
  \item \textbf{\textsc{BC-Greedy} (BC-G)} and \textbf{\textsc{BC-DualGreedy} (BC-D)} further speed up \textsc{Greedy}$^+$ and \textsc{DualGreedy}$^+$ by integrating BC-Tree, as detailed in Section \ref{sect:bctree}. They run in the same $O(ndk)$ time but exhibit improved practical efficiency. The leaf size $N_0$ of BC-Tree is set to $100$.
\end{itemize} 

All the above methods, except \textsc{Linear} and \textsc{IP-Greedy}, are implemented with both $f_{avg}(\cdot)$ and $f_{max}(\cdot)$ serving as the objective functions.
We denote their corresponding variants by suffices \textbf{\textsc{-Avg}} (or \textbf{\textsc{-A}}) and \textbf{\textsc{-Max}} (or \textbf{\textsc{-M}}), respectively.

\begin{table}[t]
\setlength\tabcolsep{0.6em}
\centering
\footnotesize
\caption{Statistics of ten real-world data sets used in the experiments.}
\label{table:dataset}
\vspace{-0.5em}
\resizebox{\columnwidth}{!}{%
  \begin{tabular}{lccccc} 
    \toprule
    \textbf{Data Set} & \textbf{\#Items} & \textbf{\#Ratings} & $\mu_{avg}$ & $\mu_{max}$ & \textbf{Category?} \\
    \midrule
    Netflix    & 17,770      & 100,480,507   & 0.05   & 0.001   &  No   \\
    MovieLens  & 59,047      & 25,000,095    & 0.05   & 0.001   &  Yes  \\
    Yelp       & 150,346     & 6,990,280     & 0.05   & 0.001   &  Yes  \\
    Food       & 160,901     & 698,901       & 0.05   & 0.001   &  Unused \\
    Google-CA  & 513,131     & 70,082,062    & 0.005  & 0.0005  &  Yes  \\
    Yahoo!     & 624,961     & 252,800,275   & 0.001  & 0.00005 &  Yes  \\
    Taobao     & 638,962     & 2,015,839     & 0.05   & 0.001   &  Unused \\
    AMZ-CSJ    & 2,681,297   & 32,292,099    & 0.05   & 0.001   &  No  \\
    AMZ-Books  & 2,930,451   & 51,311,621    & 0.05   & 0.001   &  No  \\
    LFM        & 31,634,450  & 1,088,161,692 & 0.05   & 0.001   &  No  \\
    \bottomrule
  \end{tabular}
}
\end{table}

\begin{table*}[t]
\centering
\footnotesize
\setlength\tabcolsep{4pt}
\caption{Recommendation performance of each method in terms of PCC, Cov, and Query Time (in ms) when $k = 10$.
(Note: The bold font signifies the best result for each measure; \textsc{G}/\textsc{G}$^+$ and \textsc{D}/\textsc{D}$^+$ are omitted as they have identical results to \textsc{BC-G} and \textsc{BC-D}).}
\vspace{-0.5em}
\label{tab:recommendation}
\resizebox{\textwidth}{!}{%
\begin{tabular}{llccccccccccccccccccccc}
\toprule
\multicolumn{2}{c}{\textbf{Data Set}} & \multicolumn{5}{c}{\textbf{MovieLens}} & \multicolumn{5}{c}{\textbf{Yelp}} & \multicolumn{5}{c}{\textbf{Google-CA}} & \multicolumn{5}{c}{\textbf{Yahoo!}} \\
\cmidrule(lr){1-2} 
\cmidrule(lr){3-7} \cmidrule(lr){8-12} \cmidrule(lr){13-17} \cmidrule(lr){18-22}
\multicolumn{2}{c}{$\lambda$} & 0.1 & 0.3 & 0.5 & 0.7 & 0.9 & 0.1 & 0.3 & 0.5 & 0.7 & 0.9 & 0.1 & 0.3 & 0.5 & 0.7 & 0.9 & 0.1 & 0.3 & 0.5 & 0.7 & 0.9 \\
\midrule
\multirow{6}{*}{\textbf{PCC$\uparrow$}}
& LIN & 0.806 & 0.806 & 0.806 & 0.806 & 0.806 & 0.387 & 0.387 & 0.387 & 0.387 & 0.387 & 0.071 & 0.071 & 0.071 & 0.071 & 0.071 & \textbf{0.695} & 0.695 & 0.695 & 0.695 & 0.695 \\
& IP-G & 0.594 & 0.793 & 0.806 & 0.805 & 0.805 & 0.251 & 0.257 & 0.262 & 0.278 & 0.306 & 0.028 & 0.028 & 0.033 & 0.048 & 0.058 & 0.385 & 0.438 & 0.525 & 0.575 & 0.632 \\
& BC-G-A & 0.802 & 0.807 & 0.817 & 0.827 & 0.824 & 0.391 & 0.399 & 0.401 & 0.400 & 0.402 & 0.071 & \textbf{0.082} & 0.081 & 0.078 & 0.072 & 0.628 & 0.688 & 0.710 & 0.710 & 0.710 \\
& BC-G-M & \textbf{0.814} & \textbf{0.820} & 0.818 & 0.809 & 0.828 & 0.395 & \textbf{0.403} & 0.400 & \textbf{0.401} & 0.393 & \textbf{0.076} & 0.080 & \textbf{0.085} & \textbf{0.079} & \textbf{0.073} & 0.661 & 0.710 & \textbf{0.716} & 0.708 & 0.695 \\
& BC-D-A & 0.796 & 0.806 & \textbf{0.829} & \textbf{0.829} & \textbf{0.833} & 0.390 & 0.397 & \textbf{0.402} & 0.395 & \textbf{0.404} & 0.066 & 0.071 & 0.072 & 0.074 & 0.068 & 0.648 & 0.694 & 0.707 & \textbf{0.726} & \textbf{0.731} \\
& BC-D-M & 0.810 & \textbf{0.820} & 0.820 & 0.822 & \textbf{0.833} & \textbf{0.397} & 0.402 & \textbf{0.402} & 0.398 & 0.382 & 0.072 & 0.072 & 0.080 & \textbf{0.079} & 0.067 & 0.678 & \textbf{0.713} & \textbf{0.716} & 0.705 & 0.699 \\
\midrule
\multirow{6}{*}{\textbf{Cov$\uparrow$}}
& LIN & 0.682 & 0.682 & 0.682 & 0.682 & 0.682 & 0.454 & 0.454 & 0.454 & 0.454 & 0.454 & \textbf{0.193} & 0.193 & 0.193 & 0.193 & 0.193 & 0.392 & 0.392 & 0.392 & 0.392 & 0.392 \\
& IP-G & 0.480 & 0.680 & 0.684 & 0.682 & 0.682 & 0.359 & 0.361 & 0.363 & 0.375 & 0.397 & 0.075 & 0.075 & 0.088 & 0.107 & 0.140 & 0.268 & 0.272 & 0.291 & 0.314 & 0.345 \\
& BC-G-A & \textbf{0.760} & 0.753 & 0.752 & 0.743 & \textbf{0.741} & 0.484 & 0.480 & 0.489 & 0.488 & 0.477 & 0.175 & \textbf{0.209} & 0.208 & 0.213 & 0.203 & 0.457 & \textbf{0.474} & \textbf{0.468} & \textbf{0.448} & 0.412 \\
& BC-G-M & 0.749 & 0.751 & 0.747 & 0.729 & 0.710 & 0.477 & 0.485 & 0.484 & \textbf{0.489} & 0.471 & 0.185 & 0.199 & 0.206 & 0.209 & 0.203 & 0.441 & 0.433 & 0.433 & 0.400 & 0.394 \\
& BC-D-A & 0.743 & \textbf{0.761} & \textbf{0.765} & \textbf{0.748} & 0.736 & \textbf{0.493} & \textbf{0.491} & \textbf{0.501} & 0.487 & \textbf{0.496} & 0.185 & 0.186 & 0.205 & \textbf{0.223} & \textbf{0.205} & \textbf{0.468} & 0.452 & 0.451 & 0.444 & \textbf{0.430} \\
& BC-D-M & 0.758 & 0.752 & 0.742 & 0.734 & 0.713 & 0.472 & 0.476 & 0.473 & 0.463 & 0.451 & 0.187 & 0.197 & \textbf{0.216} & 0.218 & 0.203 & 0.447 & 0.425 & 0.421 & 0.396 & 0.394 \\
\midrule
\multirow{6}{*}{\textbf{Time$\downarrow$}}
& LIN & 3.34 & 3.36 & 3.41 & 3.43 & 3.41 & \textbf{8.55} & \textbf{8.49} & \textbf{8.81} & \textbf{8.91} & 8.78 & \textbf{31.1} & \textbf{29.3} & \textbf{28.9} & \textbf{28.9} & 31.2 & \textbf{30.8} & 30.9 & 30.8 & 30.9 & 30.8 \\
& IP-G & 59.9 & 10.3 & 7.38 & 6.66 & 6.79 & 201 & 191 & 178 & 174 & 158 & 713 & 716 & 688 & 641 & 597 & 714 & 594 & 465 & 365 & 215 \\
& BC-G-A & 5.78 & 4.33 & 3.29 & 2.12 & 0.655 & 34.5 & 26.3 & 21.8 & 21.4 & 15.0 & 101 & 63.9 & 50.0 & 47.0 & 35.9 & 193 & 110 & 69.2 & 32.0 & 13.1 \\
& BC-G-M & \textbf{1.00} & \textbf{0.643} & \textbf{0.520} & \textbf{0.461} & \textbf{0.399} & 22.0 & 16.3 & 12.7 & 10.1 & \textbf{6.48} & 61.0 & 42.2 & 35.0 & 30.4 & \textbf{17.2} & 53.8 & \textbf{29.0} & \textbf{11.6} & \textbf{4.53} & \textbf{3.15} \\
& BC-D-A & 22.9 & 13.4 & 8.55 & 4.68 & 1.62 & 91.2 & 74.6 & 68.8 & 53.4 & 42.4 & 256 & 180 & 149 & 131 & 74.2 & 410 & 266 & 164 & 71.3 & 32.0 \\
& BC-D-M & 2.79 & 1.74 & 1.35 & 1.12 & 0.931 & 53.4 & 38.3 & 33.7 & 22.8 & 12.8 & 154 & 102 & 74.0 & 61.1 & 34.0 & 123 & 58.5 & 37.9 & 20.4 & 10.2 \\
\bottomrule
\end{tabular}
}
\vspace{-1.0em}
\end{table*}

\paragraph{Evaluation Measures}
As different methods employ different objective functions for D$k$MIPS, we first use two end-to-end measures to evaluate the quality of recommendations.
\begin{itemize}[nolistsep]
  \item \textbf{Pearson Correlation Coefficient (PCC)} gauges the correlation between the category (or genre) histograms of all user-rated items and the recommended items, denoted as vectors $\bar{\bm{q}}$ and $\bar{\bm{p}}$.
  Each dimension of $\bar{\bm{q}}$ and $\bar{\bm{p}}$ corresponds to a business category on Yelp and Google-CA or a genre on MovieLens and Yahoo!, i.e., $\bar{q}_i = \sum_{\bm{p} \in \mathcal{U}} r_{\bm{q}, \bm{p}} \cdot c_{\bm{p}, i}$ and $\bar{p}_i = \sum_{\bm{p} \in \mathcal{S}} c_{\bm{p}, i}$, where $r_{\bm{q}, \bm{p}}$ is the rating score of user $\bm{q}$ on item $\bm{p}$, and the term $c_{\bm{p}, i}$ serves as a binary indicator to specify whether $\bm{p}$ belongs to the category (or genre) $i$. Moreover, $\mathcal{S}$ is the set of $k$ recommended items, and $\mathcal{U}$ encompasses all items rated by $\bm{q}$.
  A higher PCC indicates a better alignment between the recommended items and the user's overall preferences.
  \item \textbf{Coverage (Cov)} quantifies the diversity of categories (or genres) within the set $\mathcal{S}$ of recommended items compared to those of all items $\mathcal{U}$ rated by user $\bm{q}$.
  It is defined as $\text{Cov}(\mathcal{S}) := \num{\textstyle \bigcup_{\bm{p} \in \mathcal{S}} C(\bm{p}) \cap \bigcup_{\bm{p} \in \mathcal{U}} C(\bm{p})} / \num{\bigcup_{\bm{p} \in \mathcal{U}} C(\bm{p})}$, where $C(\bm{p})$ is the set of categories (or genres) to which the item $\bm{p}$ belongs.
  A higher Cov means that the recommended items cover a wider range of user interests.
\end{itemize}

In addition to the two end-to-end measures, we also use the following measures to evaluate the query performance.
\begin{itemize}[nolistsep]
  \item \textbf{MMR} is computed using $f_{avg}(\mathcal{S})$ (or $f_{max}(\mathcal{S})$) as defined in Eq.~\ref{eqn:f_S-avg} (or \ref{eqn:f_S-max}). A higher MMR signifies a more optimal result for D$k$MIPS w.r.t.~the objective function.
  \item \textbf{Query Time} is the wall clock time of a method to perform a D$k$MIPS query, indicating its time efficiency.
  \item \textbf{Indexing Time} and \textbf{Index Size} measure the time and memory required for BC-Tree construction, assessing the indexing cost of \textsc{BC-Greedy} and \textsc{BC-DualGreedy}.
\end{itemize}

We repeat each experiment ten times using different random seeds and report the average results for each measure.

\begin{figure*}[t]
\centering
\includegraphics[width=0.99\textwidth]{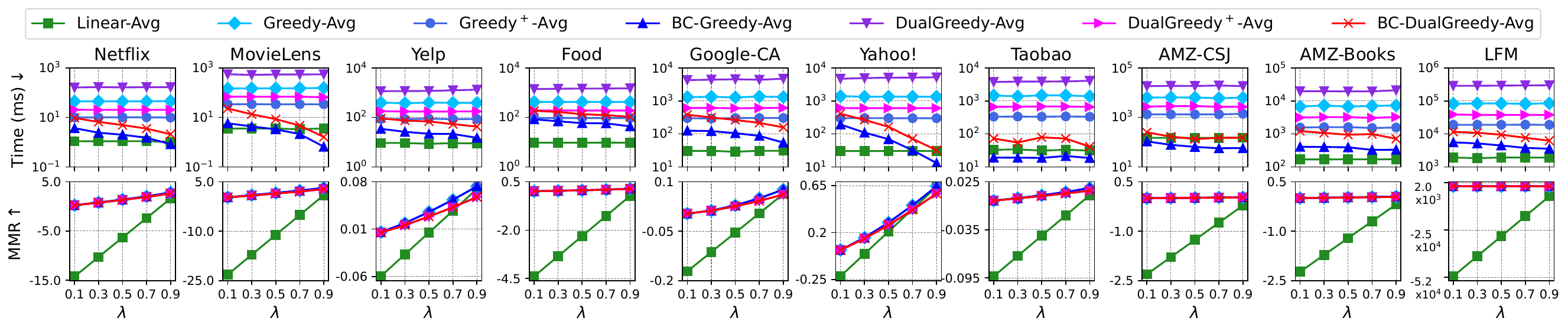}
\vspace{-1.25em}
\caption{Query performance vs.~$\lambda$ for the objective function $f_{avg}(\cdot)$ ($k=10$).}
\label{fig:query-lambda-avg}
\vspace{-0.75em}
\end{figure*}

\begin{figure*}[t]
\centering
\includegraphics[width=0.99\textwidth]{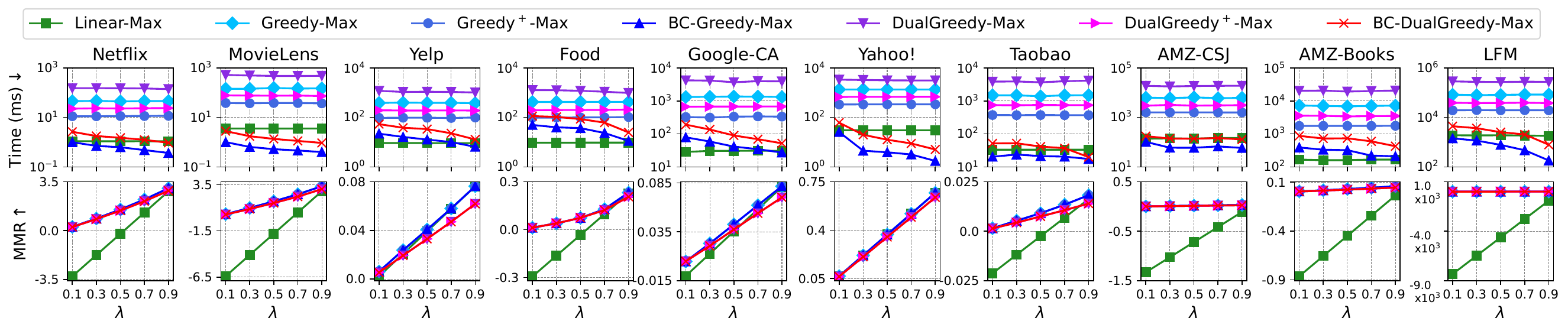}
\vspace{-1.25em}
\caption{Query performance vs.~$\lambda$ for the objective function $f_{max}(\cdot)$ ($k=10$).}
\label{fig:query-lambda-max}
\vspace{-0.75em}
\end{figure*}

\begin{figure*}[t]
\centering
\includegraphics[width=0.99\textwidth]{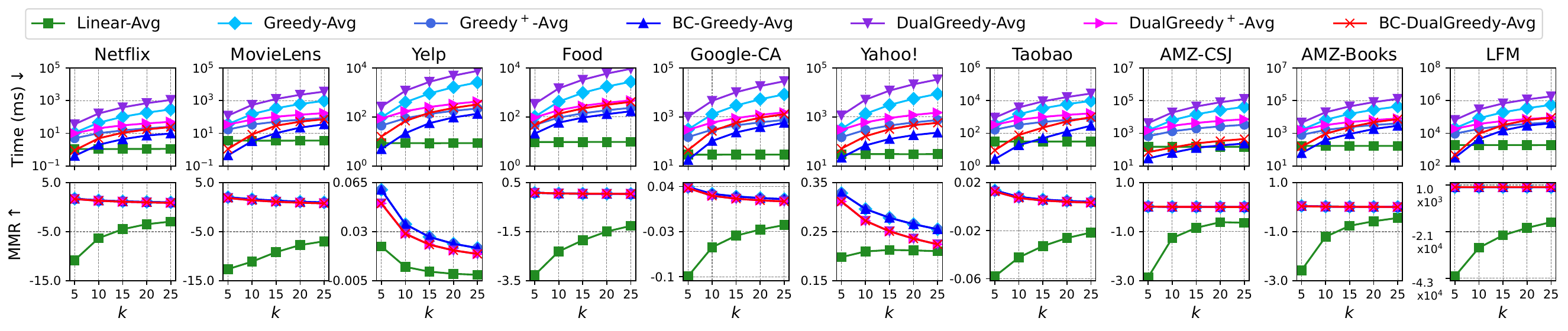}
\vspace{-1.25em}
\caption{Query performance vs.~$k$ for the objective function $f_{avg}(\cdot)$ ($\lambda=0.5$).}
\label{fig:query-k-avg}
\vspace{-0.75em}
\end{figure*}

\begin{figure*}[t]
\centering
\includegraphics[width=0.99\textwidth]{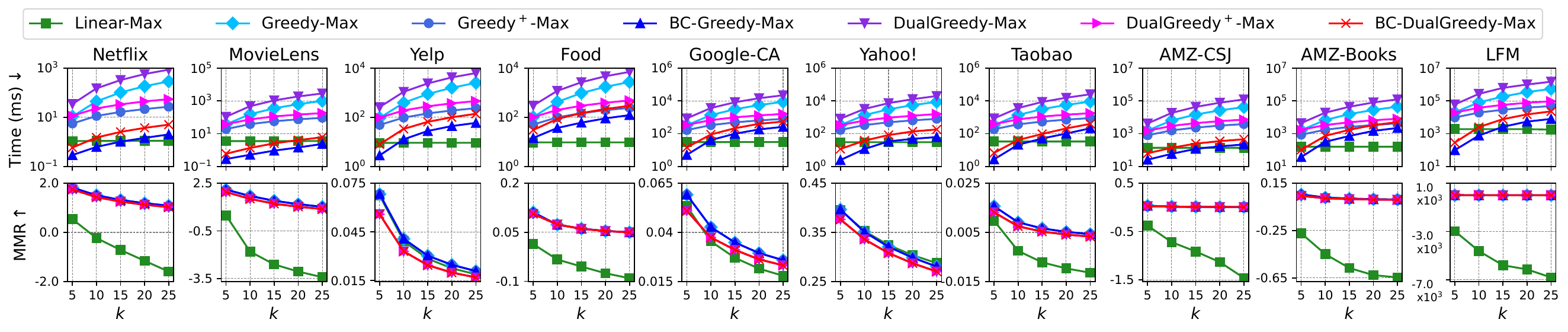}
\vspace{-1.25em}
\caption{Query performance vs.~$k$ for the objective function $f_{max}(\cdot)$ ($\lambda=0.5$).}
\label{fig:query-k-max}
\vspace{-0.75em}
\end{figure*}

\paragraph{Implementation and Environment}
We focus on in-memory workloads. All methods were written in C++ and compiled with g++-8 using -O3 optimization.
All experiments were conducted in a single thread on a Microsoft Azure cloud virtual machine with an AMD EPYC 7763v CPU @3.5GHz and 64GB memory, running on Ubuntu Server 20.04.

\paragraph{Parameter Setting}
For D$k$MIPS, we tested $k \in \{5, 10, 15,$ $20, 25\}$ and $\lambda \in \{0.1, 0.3, 0.5, 0.7, 0.9\}$.
The default $\mu$ values in ten data sets are reported in Table \ref{table:dataset}, where $\mu_{avg}$ and $\mu_{max}$ are used for $f_{avg}(\cdot)$ and $f_{max}(\cdot)$, respectively.

\begin{table*}[t]
\centering
\footnotesize
\caption{Indexing time and index size of the BC-Tree on ten real-world data sets.}
\label{tab:index}
\vspace{-0.5em}
\begin{tabular}{ccccccccccc} \toprule
  \textbf{Data Set} & Netflix  & MovieLens  & Yelp  & Food  & Google-CA  & Yahoo!  & Taobao  & AMZ-CSJ  & AMZ-Books  & LFM  \\
  \midrule
  \textbf{Indexing Time (Seconds)} & 0.313  & 1.430  & 7.516  & 3.606  & 29.42  & 15.34  & 44.80  & 246.81  & 240.29  & 3,517.75 \\
  \textbf{Index Size (MB)}         & 0.753  & 2.376  & 6.126  & 10.21  & 20.67  & 19.09  & 25.50  & 103.03  & 118.15  & 1,340.76 \\
  \bottomrule
\end{tabular}
\vspace{-0.75em}
\end{table*}


\begin{figure*}[t]%
\centering%
\subfigure[Top-10 recommendation list.]{%
  \label{fig:case-study-movielens:posters}%
  \includegraphics[width=0.58\textwidth]{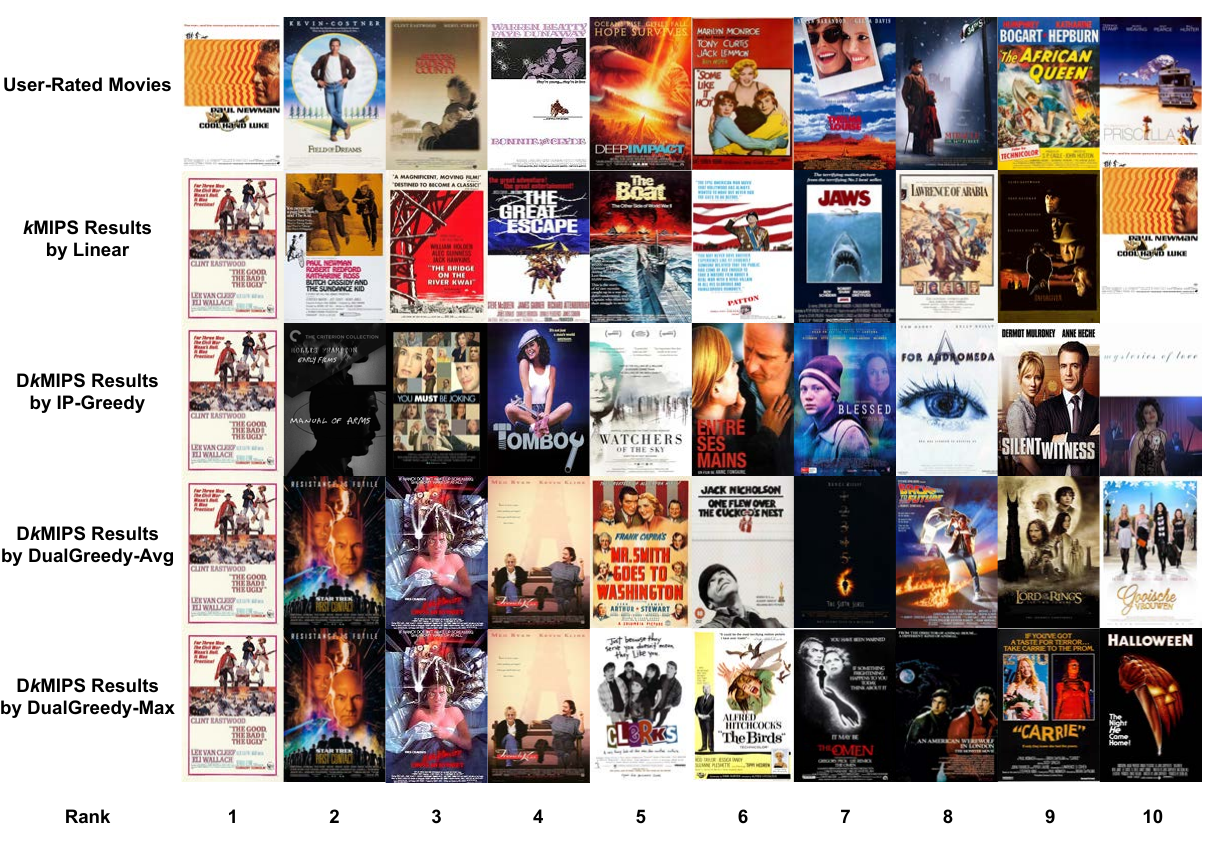}}%
\hspace{0.3em}%
\subfigure[Histogram for genre frequencies.]{%
  \label{fig:case-study-movielens:hist}%
  \includegraphics[width=0.388\textwidth]{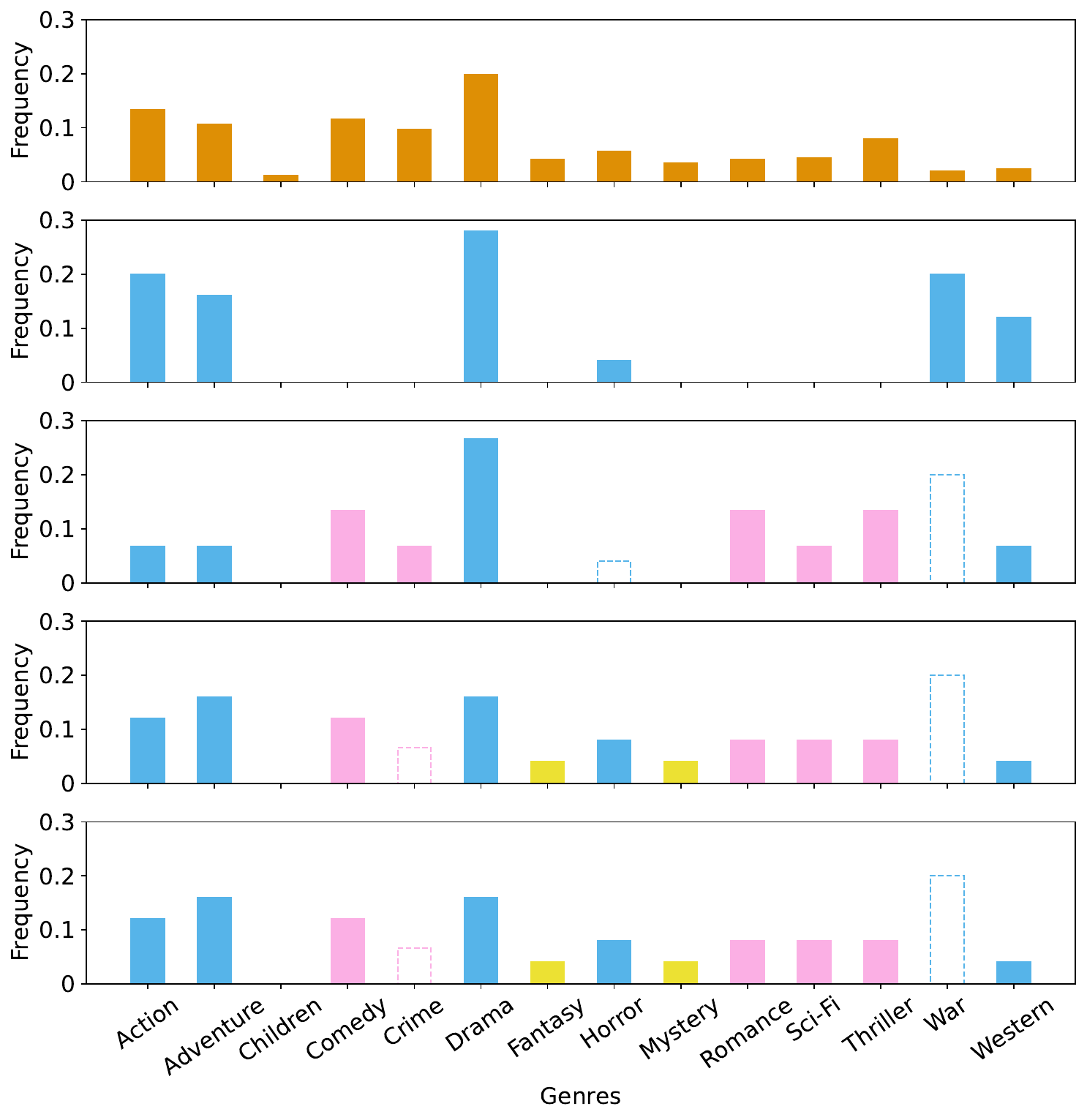}}%
\vspace{-0.5em}%
\caption{Case study on the MovieLens data set for user \#90815 ($k=10$ and $\lambda=0.5$).}%
\label{fig:case-study-movielens}%
\vspace{-0.75em}%
\end{figure*}%

\subsection{Recommendation Performance}
\label{sect:expt:recommendation}

\noindent\textbf{Recommendation Quality.}
The first two blocks of Table \ref{tab:recommendation} showcase the results for recommendation quality measured by PCC and Cov.
It is evident that our proposed methods generally achieve the highest quality across different values of $\lambda$ on all four data sets.
And \textsc{LIN} ranks second, while \textsc{IP-G} performs the worst.
This is mainly due to the misalignment of the marginal gain function of \textsc{IP-G} with its objective function, as discussed in Section \ref{sect:problem}, and its lack of theoretical guarantee.
On the contrary, the leading performance of our methods can be attributed to the revised D$k$MIPS formulation, which effectively addresses this misalignment.
Additionally, our methods come with theoretical guarantees to ensure a more balanced emphasis on relevance and diversity.
Moreover, \textsc{BC-D} surpasses \textsc{BC-G} in most cases, particularly when $f_{avg}(\cdot)$ is used. 
This is because \textsc{Greedy}-based methods offer only data-dependent approximation bound, whereas \textsc{DualGreedy}-based methods, when using $f_{avg}(\cdot)$, achieve a much better approximation ratio, as outlined in Section \ref{sect:methods:dual-greedy}.

\paragraph{Time Efficiency}
From the third block of Table \ref{tab:recommendation}, \textsc{LIN} and our proposed methods, especially \textsc{BC-G-M}, run much faster than \textsc{IP-G} by one to two orders of magnitude. 
This significant difference can be attributed to the fact that \textsc{IP-G} scales superlinearly with $n$ and quadratically with $k$ for its $O(ndk^2 \log n)$ time complexity, while \textsc{LIN} and our methods have linear time complexity w.r.t.~$n$, and \textsc{LIN} is not affected by $k$.
Upon detailed analysis, \textsc{BC-G} is at least twice as fast as \textsc{BC-D}. This is because \textsc{DualGreedy}-based methods require maintaining two separate result sets for D$k$MIPS, while \textsc{Greedy}-based methods only need one.
Furthermore, using $f_{max}(\cdot)$ is shown to be more time-efficient than using $f_{avg}(\cdot)$.
This increased efficiency is due to the typically larger value of $\Delta_{f_{max}}(\bm{p},\mathcal{S})$ compared to $\Delta_{f_{avg}}(\bm{p},\mathcal{S})$, leading to tighter upper bounds as detailed in Section \ref{sect:bctree:bounds} and thereby boosting the query efficiency.

\subsection{Query Performance}
\label{sect:expt:query}

Next, we examine the query performance of \textsc{BC-Greedy}, \textsc{BC-DualGreedy}, and \textsc{Linear} with varying $\lambda$ and $k$.

\paragraph{Performance with Varying $\lambda$}
Figs.~\ref{fig:query-lambda-avg} and \ref{fig:query-lambda-max} present the query time and MMR for $f_{avg}(\cdot)$ and $f_{max}(\cdot)$, respectively, with varying $\lambda$ from 0.1 to 0.9.
The first row of Figs.~\ref{fig:query-lambda-avg} and \ref{fig:query-lambda-max} illustrates that \textsc{BC-Greedy} and \textsc{BC-DualGreedy} can match the efficiency of \textsc{Linear}, particularly at larger values of $\lambda$. This demonstrates their practical query efficiency, as elaborated in Section \ref{sect:bctree:search}.
Although \textsc{Linear} is efficient, it tends to compromise on recommendation quality due to its lack of emphasis on \emph{diversity}.
The second row indicates that the MMRs of \textsc{BC-Greedy} and \textsc{BC-DualGreedy} consistently exceed those of \textsc{Linear} on all values of $\lambda$, benefiting from their focus on diversity.
When the value of $\lambda$ decreases, which corresponds to a case where diversity is more emphasized, they show more advantages in MMR compared to \textsc{Linear}.
These findings, along with the results of Table \ref{tab:recommendation}, confirm that our methods allow users to effectively control $\lambda$ to strike a desirable balance between relevance and diversity.

\paragraph{Performance with Varying $k$}
Figs.~\ref{fig:query-k-avg} and \ref{fig:query-k-max} show the query time and MMR for $f_{avg}(\cdot)$ and $f_{max}(\cdot)$ by ranging the result size $k$ from 5 to 25, respectively.
We find that the query time of \textsc{BC-Greedy} and \textsc{BC-DualGreedy} increases linearly with $k$, while that of \textsc{Linear} is barely affected by $k$.
The MMRs of all our methods decrease with $k$, regardless of whether $f_{avg}(\cdot)$ or $f_{max}(\cdot)$ is used.
This decrease is inevitable because the diversity term occupies a greater part of the objective function when $k$ increases.
Still, \textsc{BC-Greedy} and \textsc{BC-DualGreedy} uniformly outperform \textsc{Linear} for all $k$ values.

\subsection{Indexing Overhead}
\label{sect:expt:indexing}

Table \ref{tab:index} presents the indexing overhead of BC-Tree on ten data sets.
The construction time and memory usage of BC-Tree are nearly linear with $n$, as have been analyzed in Theorem \ref{theorem:ball_tree_construction}. 
Notably, for the largest data set, LFM, with around 32 million items, the BC-Tree is constructed within an hour and takes about 1.3GB of memory.
As the index is built only once and can be updated efficiently, its construction is lightweight with modern hardware capabilities.

\subsection{Case Study}
\label{sect:expt:case_study}

Fig.~\ref{fig:case-study-movielens} shows a case study conducted on the MovieLens data set.
From the first row, user \#90815 exhibits diverse interests in many movie genres, including but not limited to \emph{Drama, Action, Adventure, Comedy, Thriller}, etc.
\textsc{Linear}, as depicted in the second row, returns movies primarily from the most relevant genres to the user, such as \emph{Action, Adventure, Drama, War, and Western}. The corresponding posters also reveal a lack of diversity, displaying predominantly oppressive styles.
The third row indicates that the \textsc{IP-Greedy} result has a more diverse genre distribution. From its RHS histogram, \textsc{IP-Greedy} includes movies from five additional genres compared to \textsc{Linear} while missing two genres that \textsc{Linear} has recommended.
\textsc{DualGreedy-Avg} and \textsc{DualGreedy-Max}, as shown in the last two rows, further extend the diversity of genres by including three extra genres compared to \textsc{IP-Greedy}, albeit missing one (less relevant) genre discovered by \textsc{IP-Greedy}.
The posters and histograms affirm that the \textsc{DualGreedy}-based methods provide a more diverse and visually appealing movie selection than the baselines.

\begin{table*}[t]
\centering
\footnotesize
\setlength\tabcolsep{4pt}
\caption{Recommendation quality (PCC \& Cov) results for \textsc{BC-D-A} and \textsc{BC-D-M} with or without the item representation by Item2Vec (I2V).}
\vspace{-0.5em}
\label{tab:twospace}
\resizebox{\textwidth}{!}{%
  \begin{tabular}{llccccccccccccccccccccc}\toprule
    \multicolumn{2}{c}{\textbf{Data Set}} & \multicolumn{5}{c}{\textbf{MovieLens}} & \multicolumn{5}{c}{\textbf{Yelp}} & \multicolumn{5}{c}{\textbf{Google-CA}} & \multicolumn{5}{c}{\textbf{Yahoo!}} \\
    \cmidrule(lr){1-2} \cmidrule(lr){3-7} \cmidrule(lr){8-12} \cmidrule(lr){13-17} \cmidrule(lr){18-22}
    \multicolumn{2}{c}{$\lambda$} & 0.1 & 0.3 & 0.5 & 0.7 & 0.9 & 0.1 & 0.3 & 0.5 & 0.7 & 0.9 & 0.1 & 0.3 & 0.5 & 0.7 & 0.9 & 0.1 & 0.3 & 0.5 & 0.7 & 0.9 \\
    \midrule
    \multirow{5}{*}{\textbf{PCC$\uparrow$}}
    & \textsc{IP-G} & 0.594 & 0.793 & 0.806 & 0.805 & 0.805 & 0.251 & 0.257 & 0.262 & 0.278 & 0.306 & 0.028 & 0.028 & 0.033 & 0.048 & 0.058 & 0.385 & 0.438 & 0.525 & 0.575 & 0.632 \\
    & \textsc{BC-D-A w I2V} & 0.789 & 0.814 & 0.809 & 0.809 & 0.807 & 0.384 & 0.397 & 0.397 & 0.394 & 0.389 & 0.061 & \textbf{0.080} & 0.080 & 0.076 & 0.074 & 0.689 & 0.693 & 0.699 & 0.697 & 0.695 \\
    & \textsc{BC-D-M w I2V} & 0.778 & 0.810 & 0.808 & 0.808 & 0.806 & 0.357 & 0.390 & 0.388 & 0.391 & 0.387 & 0.060 & 0.074 & \textbf{0.084} & \textbf{0.080} & \textbf{0.079} & \textbf{0.694} & 0.696 & 0.695 & 0.695 & 0.695 \\
    & \textsc{BC-D-A w/o I2V} & 0.796 & 0.806 & \textbf{0.829} & \textbf{0.829} & \textbf{0.833} & 0.390 & 0.397 & \textbf{0.402} & 0.395 & \textbf{0.404} & 0.066 & 0.071 & 0.072 & 0.074 & 0.068 & 0.648 & 0.694 & 0.707 & \textbf{0.726} & \textbf{0.731} \\
    & \textsc{BC-D-M w/o I2V} & \textbf{0.810} & \textbf{0.820} & 0.820 & 0.822 & \textbf{0.833} & \textbf{0.397} & \textbf{0.402} & \textbf{0.402} & \textbf{0.398} & 0.382 & \textbf{0.072} & 0.072 & 0.080 & 0.079 & 0.067 & 0.678 & \textbf{0.713} & \textbf{0.716} & 0.705 & 0.699 \\
    \midrule
    \multirow{5}{*}{\textbf{Cov$\uparrow$}}
    & \textsc{IP-G} & 0.480 & 0.680 & 0.684 & 0.682 & 0.682 & 0.359 & 0.361 & 0.363 & 0.375 & 0.397 & 0.075 & 0.075 & 0.088 & 0.107 & 0.140 & 0.268 & 0.272 & 0.291 & 0.314 & 0.345 \\
    & \textsc{BC-D-A w I2V} & 0.684 & 0.686 & 0.685 & 0.682 & 0.682 & 0.469 & 0.478 & 0.482 & 0.467 & 0.460 & 0.152 & \textbf{0.197} & 0.202 & 0.206 & 0.203 & 0.406 & 0.403 & 0.405 & 0.402 & 0.392 \\
    & \textsc{BC-D-M w I2V} & 0.665 & 0.682 & 0.682 & 0.682 & 0.682 & 0.462 & 0.475 & 0.472 & 0.472 & 0.457 & 0.147 & 0.190 & 0.214 & 0.209 & \textbf{0.218} & 0.403 & 0.402 & 0.392 & 0.392 & 0.392 \\
    & \textsc{BC-D-A w/o I2V} & 0.743 & \textbf{0.761} & \textbf{0.765} & \textbf{0.748} & \textbf{0.736} & \textbf{0.493} & \textbf{0.491} & \textbf{0.501} & \textbf{0.487} & \textbf{0.496} & 0.185 & 0.186 & 0.205 & \textbf{0.223} & 0.205 & \textbf{0.468} & \textbf{0.452} & \textbf{0.451} & \textbf{0.444} & \textbf{0.430} \\
    & \textsc{BC-D-M w/o I2V} & \textbf{0.758} & 0.752 & 0.742 & 0.734 & 0.713 & 0.472 & 0.476 & 0.473 & 0.463 & 0.451 & \textbf{0.187} & \textbf{0.197} & \textbf{0.216} & 0.218 & 0.203 & 0.447 & 0.425 & 0.421 & 0.396 & 0.394 \\
    \bottomrule
  \end{tabular}
}
\vspace{-1.0em}
\end{table*}

\subsection{Ablation Study}

\noindent\textbf{Effect of Optimization Techniques and BC-Tree.}
We first assess the effectiveness of each component in \textsc{BC-Greedy} and \textsc{BC-DualGreedy}.
From Figs.~\ref{fig:query-lambda-avg} and \ref{fig:query-lambda-max}, we find that \textsc{BC-Greedy} and \textsc{BC-DualGreedy} run much faster than \textsc{Greedy}$^+$ and \textsc{DualGreedy}$^+$. And they are consistently faster than \textsc{Greedy} and \textsc{DualGreedy}.
This validates the efficiency of our optimization techniques outlined in Section \ref{sect:methods:optimizations} and the BC-Tree integration explored in Section \ref{sect:bctree:search}.
Notably, the performance of all BC-Tree-based methods improves as $\lambda$ increases, demonstrating that the upper bounds derived from BC-Tree are more effective when relevance is more emphasized.
We also observe that the MMRs for all \textsc{Greedy}-based (as well as \textsc{DualGreedy}-based) methods are the same, reflecting their shared goal of speeding up D$k$MIPS without changing the results.
Similar results in Figs.~\ref{fig:query-k-avg} and \ref{fig:query-k-max} further validate the effectiveness of each component.

\paragraph{Performance with/without Item2Vec}
We then study whether the item vectors generated by Item2Vec (\textsc{I2V}) can improve the efficacy of our methods in solving D$k$MIPS. 
We adapted \textsc{BC-DualGreedy} (\textsc{BC-D}) to measure diversity using Euclidean distances between \textsc{I2V} vectors.
Table \ref{tab:twospace} reveals that \textsc{BC-D-w/o-I2V} often outperforms or matches \textsc{BC-D-w-I2V}. This supports our claim in Section \ref{sect:intro} that using two item representations derived from the same rating matrix does not benefit recommendation quality.
In addition, the superior performance of \textsc{BC-D-w-I2V} over \textsc{IP-Greedy} underscores the efficacy of our new D$k$MIPS formulation and methods.

\begin{figure}[ht]
\centering
\vspace{-0.5em}
\includegraphics[width=0.99\columnwidth]{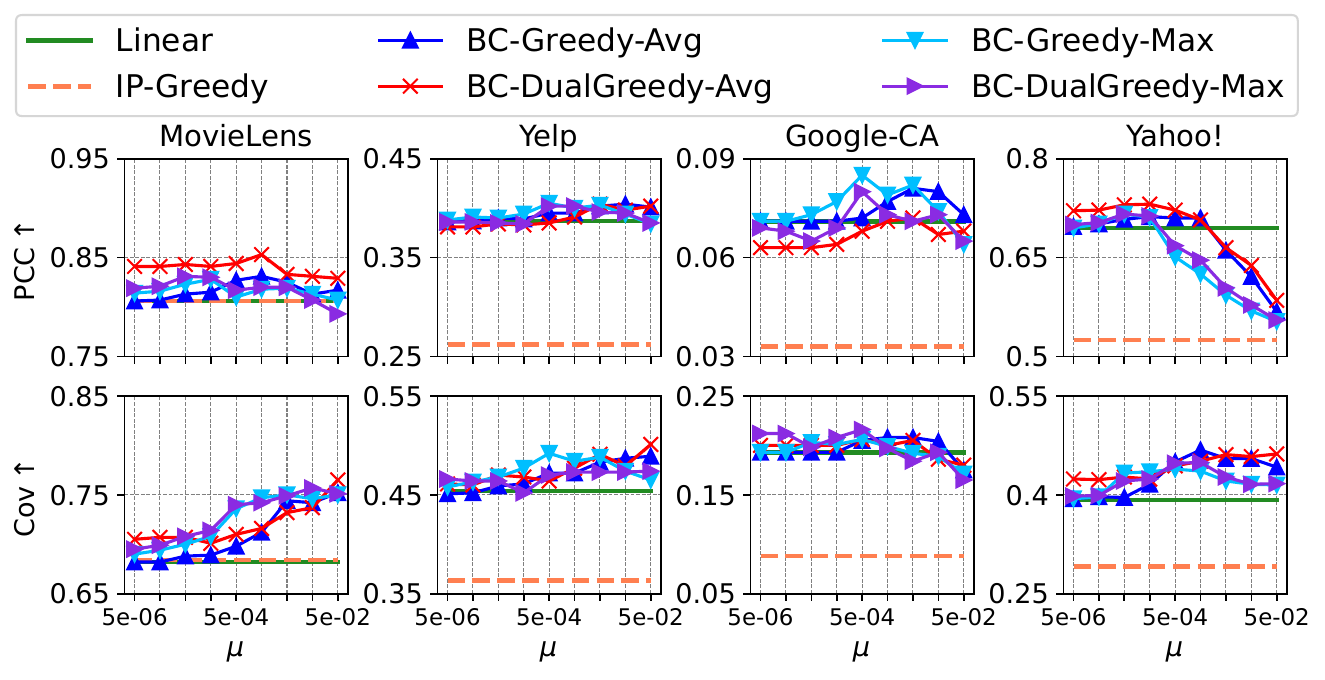}
\vspace{-2.25em}
\caption{Recommendation quality vs.~$\mu$ ($k=10$ and $\lambda=0.5$).}
\label{fig:rec_perf_mus}
\vspace{-0.75em}
\end{figure}

\subsection{Impact of Scaling Factor}
\label{sect:expt:para}

Fig.~\ref{fig:rec_perf_mus} displays the PCC and Cov of different methods by adjusting the scaling factor $\mu$ in their objective functions.
As \textsc{Linear} and \textsc{IP-Greedy} lack a tunable $\mu$, their PCC and Cov are constant, depicted as horizontal lines.
For \textsc{BC-Greedy} and \textsc{BC-DualGreedy}, we varied $\mu$ from $5 \times 10^{-6}$ to $5 \times 10^{-2}$.
A notable finding is that no single $\mu$ value is optimal because the $l_2$-norms of item vectors vary across data sets and the ranges of the values of $f_{avg}(\cdot)$ and $f_{max}(\cdot)$ are different.
However, \textsc{BC-Greedy} and \textsc{BC-DualGreedy} consistently outperform \textsc{Linear} and \textsc{IP-Greedy} in a broad range of $\mu$ for both objective functions.
This suggests that a good value of $\mu$ can be quickly identified by binary search, highlighting the robust efficacy of \textsc{BC-Greedy} and \textsc{BC-DualGreedy} in recommendation scenarios.

\begin{figure}[ht]
\centering
\vspace{-0.75em}
\includegraphics[width=0.99\columnwidth]{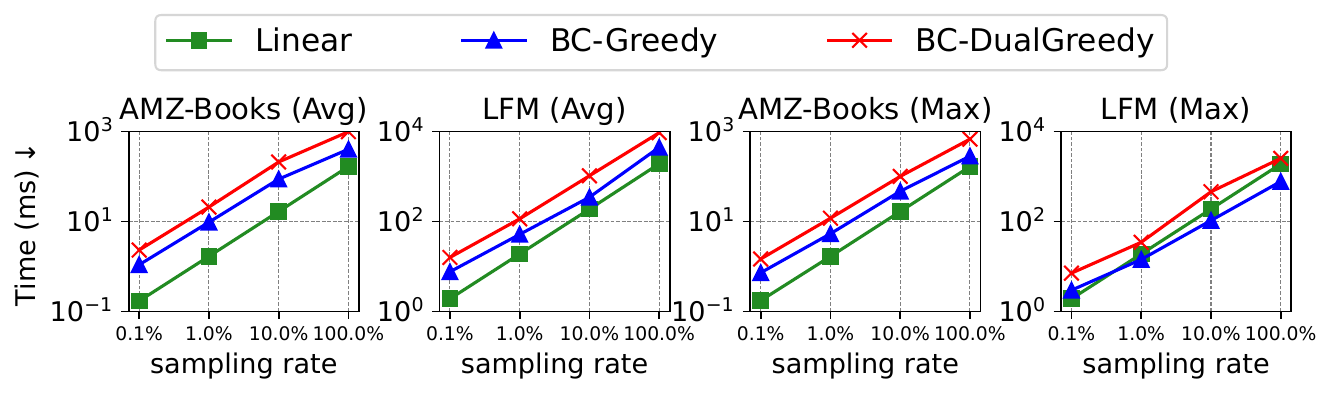}
\vspace{-2.25em}
\caption{Query time vs. sampling rate ($k=10$ and $\lambda=0.5$).}
\label{fig:scalability}
\vspace{-0.75em}
\end{figure}

\subsection{Scalability Test}
\label{sect:expt:scalability}

Finally, we examine the scalability of \textsc{BC-Greedy} and \textsc{BC-DualGreedy} w.r.t.~$n$ by adjusting the number of items through randomized sampling. 
Fig.~\ref{fig:scalability} reveals that they exhibit a near-linear growth in query time as $n$ increases.
A 1,000-fold increase in the number of items leads to a similar increase in the query time, affirming the strong scalability of BC-Tree-based methods for handling large data sets.

\section{Related Work}
\label{sect:related_work}

\noindent\textbf{$k$-Maximum Inner Product Search ($k$MIPS).}
The $k$MIPS problem has been extensively studied, resulting in two broad categories of methods: index-free and index-based.
Index-free methods do not use any auxiliary data structure for query processing, which can be further divided into scan-based \cite{TeflioudiGM15, TeflioudiG17, LiCYM17, AbuzaidSBZ19} and sampling-based approaches \cite{BallardKPS15, YuHLD17, DingYH19, LorenzenP20, Pham21}.
Scan-based methods perform linear scans on vector arrays, boosting efficiency through strategic pruning and skipping, while sampling-based methods perform efficient sampling for approximate $k$MIPS with low (expected) time complexities.
Index-based methods, on the other hand, leverage extra data structures for $k$MIPS, including tree-based \cite{RamG12, KoenigsteinRS12, CurtinR14, KeivaniSR18}, locality-sensitive hashing (LSH) \cite{Shrivastava014, NeyshaburS15, Shrivastava015, HuangMFFT18, YanLDCC18, SongGZ021, ZhaoZYLXZJ23}, quantization-based \cite{ShenLZYS15, GuoKCS16, DaiYNLC20, xiang2021gaips, Zhang_Lian_Zhang_Wang_Chen_2023}, and proximity graph-based methods \cite{MorozovB18, TanZXL19, ZhouTX019, LiuYDLCY20, TanXZFZL21}.
Recent studies also explored $k$MIPS variants such as inner product similarity join \cite{ahle2016complexity, nakama2021approximate} and reverse $k$MIPS \cite{AmagataH21, huang2023sah, amagata2023reverse}.
Although effective for $k$MIPS and its variants, they do not typically consider result diversification. In this work, we target the D$k$MIPS problem and develop new methods to provide relevant and diverse results to users.

\paragraph{Query Result Diversification}
The diversification of query results has been a topic of enduring interest in databases and information retrieval (see \cite{ZhengWQLG17, kunaver2017diversity} for extensive surveys).
Generally, diversification aims to (1) reduce redundancy in query results, (2) address ambiguous user queries, and (3) improve the generalizability of learning-based query systems.
Existing approaches are broadly classified into in-processing and post-processing techniques. 
In-processing methods concentrate on integrating diversity into loss functions of ranking \cite{YanQPWB21} and recommendation systems \cite{ChenRCSR22}, which is interesting but out of the scope of this work.
Conversely, post-processing methods like ours incorporate a diversity term to query objective functions to produce more diverse results.
Various post-processing strategies exist for different query types, such as keyword \cite{AgrawalGHI09, GollapudiS09}, relational database \cite{QinYC12, DrosouP12}, spatio-temporal \cite{CaiKFMP20, YokoyamaH16, GuoJTZ18} and graph queries \cite{FanWW13, YuanQLCZ16, LiuJYZ18}.
To our knowledge, diversity-aware query processing in high-dimensional inner product spaces remains largely unexplored \cite{hirata2022solving, hirata2023categorical}.
In this paper, we revisit the D$k$MIPS problem and method in \cite{hirata2022solving} and address their limitations by introducing a novel D$k$MIPS formulation and several methods with significantly better recommendation performance.

\section{Conclusion}
\label{sect:conclusions}

In this paper, we investigate the problem of balancing relevance and diversity in $k$MIPS for recommendations.
We revisit and explore the D$k$MIPS problem to seamlessly integrate these two critical aspects.
We introduce \textsc{Greedy} and \textsc{DualGreedy}, two approximate linear scan-based methods for D$k$MIPS.
We further improve the efficiency of both methods through BC-Tree integration so that they are suitable for real-time recommendations.
Extensive experiments on ten real-world data sets demonstrate the superior performance of our methods over existing ones in providing relevant and diverse recommendations.
These advances highlight not only the importance of balancing relevance and diversity in recommender systems but also the necessity of incorporating application-specific factors to improve query results.


\balance
\bibliographystyle{IEEEtranS}
\bibliography{IEEEabrv,main}

\end{document}